\documentclass[aps,showpacs,superscriptaddress,nofootinbib]{revtex4}
\usepackage[dvips]{graphicx}
\usepackage{epsfig,rotating}
\usepackage{amsmath,amssymb}
\numberwithin{equation}{section}

\newcommand{\be}{\begin{equation}}
\newcommand{\ee}{\end{equation}}
\newcommand{\bea}{\begin{eqnarray}}
\newcommand{\eea}{\end{eqnarray}}
\begin{document}
\title{Photon production from a thermalized quark gluon plasma: \\
quantum kinetics and nonperturbative aspects}
\author{D. Boyanovsky}
\email{boyan@pitt.edu} \affiliation{Department of Physics and
Astronomy, University of Pittsburgh, Pittsburgh, Pennsylvania
15260, USA}\affiliation{LPTHE, Universit\'e Pierre et Marie Curie
(Paris VI) et Denis Diderot (Paris VII), Tour 16, 1er. \'etage, 4,
Place Jussieu, 75252 Paris, Cedex 05, France}
\author{H. J. de Vega}
\email{devega@lpthe.jussieu.fr} \affiliation{LPTHE, Universit\'e
Pierre et Marie Curie (Paris VI) et Denis Diderot (Paris VII),
Tour 16, 1er. \'etage, 4, Place Jussieu, 75252 Paris, Cedex 05,
France}\affiliation{Department of Physics and Astronomy,
University of Pittsburgh, Pittsburgh, Pennsylvania 15260, USA}

\date{\today}

\begin{abstract}
We study the production of photons from a quark gluon plasma in
local thermal equilibrium by introducing a non-perturbative
formulation of the real time evolution of the density matrix. The
main ingredient is the real time effective action for the
electromagnetic field to $\mathcal{O}(\alpha_{em})$ and to all
orders in $\alpha_s$. The real time evolution is completely
determined by the solution of a \emph{classical stochastic}
non-local Langevin equation which provides a Dyson-like
resummation of the perturbative expansion. The Langevin equation
is solved in closed form by Laplace transform in terms of the
thermal photon polarization. A quantum kinetic description emerges
directly from this formulation.  We find that photons with $k
\lesssim 200 ~\mbox{Mev}$ \emph{thermalize} as plasmon
quasiparticles in the plasma on time scales $t \sim 10-20
~\mbox{fm}/c$ which is of the order of the lifetime of the QGP
expected at RHIC and LHC. We then obtain the direct photon yield
to lowest order in $\alpha_{em}$ and to leading logarithmic order
in $\alpha_s$ in a \emph{uniform} expansion valid at all time. The
yield during a QGP lifetime $t \sim 10 ~\mbox{fm}/c$ is
systematically larger than that obtained with the equilibrium
formulation and the spectrum features a distinct flattening for $k
\gtrsim 2.5 ~\mbox{Gev}$. We discuss the window of reliability of
our results, the theoretical uncertainties in \emph{any} treatment
of photon emission from a QGP in LTE and the shortcomings of the
customary S-matrix approach.
\end{abstract}

 \maketitle

\section{Introduction}
The quark gluon plasma (QGP) is a novel state of matter
conjectured to be formed during ultrarelativistic heavy ion
collisions and to have existed when the Universe was about
$10~\mu\mbox{secs}$ old. Ultrarelativistic heavy ion experiments
at SPS-CERN, AGS-BNL, RHIC-BNL and the forthcoming LHC at CERN
seek to create this state in collisions of  heavy ions such as
$Pb$ and $Au$ up to $\sqrt{s}\sim 200 ~\mbox{Gev}/\mbox{nucleon}$
at RHIC and $5~\mbox{Tev}/\mbox{nucleon}$ at LHC . Current
theoretical ideas suggest that a QGP in local thermodynamic
equilibrium (LTE) is formed on a time scale of the order of $\sim
1~\mbox{fm}/c$ after the collision when the partons in the
colliding nuclei are liberated. Parton-parton collisions during a
pre-equilibrium stage is then conjectured to lead to a state of
local thermodynamic equilibrium that expands hydrodynamically and
eventually undergoes a hadronization phase transition at a
temperature of order $\sim 160 ~\mbox{Mev}$\cite{revius}. The
experimental confirmation  of a quark gluon plasma hinges on
identifying observables that are directly associated with the
properties of the QGP. Electromagnetic probes, namely photons and
lepton pairs are considered to be ``clean'' since they only
interact electromagnetically and their mean free paths are
expected to be much larger than the size of the QGP. These probes
are expected to leave the hot and dense region without further
scattering, hence carrying direct information of the
QGP\cite{feinberg,mclerran,kapusta}. These expectations led to an
effort to obtain an assessment of direct photon emission and
production of dilepton pairs from a thermalized
QGP\cite{feinberg}-\cite{renk}. Preliminary assessments concluded
that direct photon emission from a QGP in LTE can be larger than
electromagnetic emission from a hadronic gas\cite{kapusta,gale}.

The first observation of direct photon production in
ultrarelativistic heavy ion collisions has been reported by the
WA98 collaboration at SPS-CERN in
$^{208}\mbox{Pb}+^{208}\mbox{Pb}$ at
$\sqrt{s}=158~\mbox{Gev}/\mbox{nucleon}$\cite{WA98}. The WA98 data
reveals an excess of direct photons above the expected background
from hadronic decays in the range of transverse momentum $p_T >
1.5 ~\mbox{Gev}$ in the most central collisions. These
observations confirm the feasibility of direct photons as
experimental probes for studying the formation and evolution of a
QGP.

A variety of fits of the experimental data with various
theoretical models has been reported\cite{revphoton}, however the
results are inconclusive: models with or without QGP emission seem
to fit the data in a manner qualitatively not very different from
the fits based solely on hadronic `cocktails'\cite{revphoton}.

Photon data from ultrarelativistic heavy ion collisions at RHIC
are forthcoming and we believe it is imperative to re-assess the
current theoretical understanding of direct photon emission from a
QGP in LTE for a reliable theoretical prediction, upon which may
hinge the identification of a QGP.

The most widely used approach to study photon production from a
QGP in LTE is based on the S-matrix calculation of an inclusive
transition probability per unit space-time volume with initial
asymptotic states of quarks and gluons averaged with a thermal
distribution. Recent studies in real-time have provided a detailed
criticism of this approach on the basis that the S-matrix
calculation explicitly assumes an infinite lifetime for a QGP in
LTE and treats the uncofined quarks and gluons in the plasma as
initial asymptotic states\cite{nuestro,wangphoton}. The underlying
assumptions in the widely used S-matrix approach, namely, treating
the medium as of infinite lifetime and spatial extent are
manifestly inconsistent with the experimental and physical
situation in which the state prior to the collision is described
by hadrons and the ensuing QGP, if formed, only lasts during a
(proper) time of order $\sim 10 \mbox{fm}/c$ and its spatial scale
is $\sim 10 fm$. There are preliminary studies of the influence of
finite \emph{size} effects on photon emission from the
plasma\cite{wong,sarkar} and more recently for dilepton emission
from the hadronic gas\cite{ruuska2}. Studies of the finite
\emph{lifetime} effects on the photon production yield were
reported in references\cite{wangphoton,nuestro}. The results of
these studies point out the importance of non-equilibrium real
time processes whose contribution is subleading in the limit of
infinite lifetime but that during the \emph{finite} lifetime yield
contributions of the same order of or larger than those found in
the S-matrix approach.

Studying particle emission from a QGP is fundamentally different
from a scattering experiment. In a scattering experiment the beam
consists of physical particles, namely asymptotic ``in'' states,
while in a QGP the quarks and gluons exist as a deconfined state
of matter as a \emph{transient} state. The ``in''and ``out''
states are hadrons not quarks. The QGP thus emerges as an
intermediate, transient state, and treating it as stationary
source of electromagnetic radiation as is explicit in the S-matrix
approach should be taken, at best, as an approximation.

The fundamentally correct calculation of photon production in an
ultrarelativistic heavy ion collision must necessarily  begin from
an in state of nuclei formed by confined quark and gluons. These
bound states evolve from the infinite past with the full
Hamiltonian continuing the evolution through the collision and
deconfinement process, through the thermalization and formation of
a thermal QGP, through hadronization and eventually through final
state interactions after freeze out. The photons produced during
all of these processes will be measured as out states in the
detector. Photons are produced at every stage, and the total
number of photons \emph{can} be calculated from the S-matrix if
\emph{all of the processes} are included in the dynamics with the
inclusive out states being hadrons. Of course this is so far an
unsolved theoretical task, and the problem emerges when one
attempts to isolate individual stages of the dynamics and treat
these stages, each of which has a finite time duration, in terms
of an S-matrix calculation. For example, current estimates neglect
the photons produced prior to the collision, treats the
pre-equilibrium stage via parton cascade models, and ignores the
photons from this stage in the calculation of emission from a QGP,
which itself is treated as a steady state\cite{nuestro}.

The potential identification of the formation and evolution of a
QGP from electromagnetic probes from forthcoming RHIC data
requires a reliable theoretical understanding of the phenomena.

In particular a recent detailed study of photon production in real
and finite time from a QGP in LTE reveals that the spectrum of
photons emitted during the finite lifetime of the QGP depends on
the initial state at the time of thermalization and of the
pre-equilibrium stage. This dependence is more pronounced for
large momenta $k \gtrsim 4-5 ~\mbox{Gev}$. Such conclusion is in
agreement with those obtained in ref.\cite{renk} wherein a
sensitivity of the large $k_T$ part of the spectrum to initial
conditions was also found. The real time studies of photon
production from a QGP in LTE with a finite lifetime in
refs.\cite{wangphoton,nuestro} all reveal a flattening of the
spectrum at large momentum, the precise value of the momentum at
which the spectrum flattens being sensitive to the pre-equilibrium
stage which determines the initial state of the plasma at the
thermalization hypersurface.

Perhaps coincidentally the WA98 data\cite{WA98} displays a
flattening of the spectrum for $p_T \gtrsim 1.5 ~\mbox{Gev}$.

\vspace{1mm}

\textbf{Window of opportunity:} Before a calculation of the direct
photon yield from a QGP in LTE is attempted, it is important to
establish the regime of experimental relevance and of theoretical
reliability  of \emph{any} prediction of the yield based on LTE.
The region of soft photons $k \lesssim 100- 200 ~\mbox{Mev}$ is
experimentally complicated by the enormous background of photons
from neutral pion decay (produced profusely in ultrarelativistic
heavy ion collisions), bremsstrahlung from final state
interactions etc. Assuming that LTE is established and maintained
by quark and gluon collisions and assuming a collisional mean free
path of order of $0.1 \lesssim \lambda \lesssim 0.3 ~\mbox{fm}$,
photons with momenta $ k \gtrsim 3-5 ~\mbox{Gev}$ will likely
probe scales shorter than the mean free path where the LTE
approximation breaks down. Large departures from hydrodynamic
behavior entailed by LTE is revealed in recent elliptic flow data
on the parameter $v_2(p_T)$\cite{v2pT,jacobs} for $p_T >
2~\mbox{Gev}$. While this data indicates the breakdown of (ideal)
hydrodynamics for hadrons, a similar breakdown is expected on
physical grounds for hard electromagnetic probes when the Compton
wavelength of the probe is smaller than the scale of the mean free
path. Thus the reliability of a calculation of hard particle
emission from a QGP in LTE must be re-assessed.

Furthermore the spectrum of high energy photons for $k \gtrsim 4-5
~\mbox{Gev}$ is sensitive to the initial conditions an the
pre-equilibrium stage\cite{renk,nuestro}. Originally the
pre-equilibrium stage was modelled as a parton
cascade\cite{geiger}, however more recently a different picture is
emerging for the description of the pre-equilibrium stage based on
color glass condensates\cite{glass}. Clearly the current
understanding of the pre-equilibrium stage is still a matter of
ongoing study.

Thus from the experimental point of view the relevant range of
``clean'' photons is probably for $k \gtrsim 100-200 ~\mbox{Mev}$,
and from a theoretical point of view, the reliability of
\emph{any} calculation of the yield based on the assumption of LTE
is probably suspect for $k \gtrsim 4-5 ~\mbox{Gev}$ with the added
uncertainty of dependence on the pre-equilibrium stage for momenta
larger than this range. Therefore, in this article we will focus
our study to the range $200 ~\mbox{Mev} \lesssim  k \lesssim 5
~\mbox{Gev}$  commenting on further uncertainties emerging from
our study for the high energy region.

\textbf{Goals of this article:} In this article we study
non-perturbative aspects of the real time dynamics of photon
production and propagation with the goal of obtaining a deeper
understanding of the direct photon yield during the finite
lifetime of a transient QGP in LTE. In particular we focus on
establishing a real time description of photon production and
obtaining a theoretical prediction of the yield and the spectrum
by including processes that are \emph{not} included in the usual
S-matrix calculation even at lowest order in $\alpha_{em}$. These
processes lead to subdominant contributions in the asymptotically
long time limit, but their contribution during the finite lifetime
of the QGP is of the same order of \textbf{or larger than} those
extracted solely from an S-matrix analysis.

We begin by obtaining the time evolution of an initial density
matrix directly from the effective action for the electromagnetic
field \emph{exact} to order $\alpha_{em}$ and to \emph{all} orders
in the strong coupling $\alpha_s$.  This formulation makes
manifest the connection between photon production and the
stochastic nature of photon emission and propagation in a
thermalized plasma. The expression for the photon production yield
obtained from this description reproduces the S-matrix results in
a strict perturbative expansion in $\alpha_{em}$ for a steady
state QGP. Moreover, when the full evolution is taken into account
this formulation includes the dynamics of photon propagation in
the medium and of thermalization. Furthermore, this formulation
reproduces the results of kinetic theory, highlighting the
limitations of the equilibrium approach to photon production.

\textbf{Summary of main results:}
\begin{itemize}

\item{We provide a formulation of direct photon production in real
time by obtaining the effective action for the electromagnetic
field  up to lowest order in $\alpha_{em}$ and in principle to all
orders in $\alpha_s$. Integrating out the quark and gluon fields
to obtain the effective action leads to the description of the
production and propagation of photons in a thermal bath. The
properties of the QGP thermal bath are determined by
current-current correlation functions and a stochastic gaussian
colored noise that obey a generalized fluctuation-dissipation
relation completely determined by the thermal photon polarization.
The time evolution of the photon distribution function is
determined by the solution of a \emph{classical stochastic and
non-local} Langevin equation. We explicitly solve the non-local
Langevin equation in closed form by Laplace transform in terms of
the thermal photon polarization. The resulting evolution
represents a Dyson-like resummation of the naive perturbative
expansion and provides a \textbf{uniform} expansion in
$\alpha_{em}$ valid at all times and which includes
non-perturbative aspects. We obtain the photon yield and the
spectrum during the lifetime of the QGP expected at RHIC ($\approx
10~\mbox{fm}/c$) up to leading order in $\alpha_{em}$ and to
logarithmic order in $\alpha_s$.}

 \item{An important aspect that
emerges from our study is that intermediate energy photons with $k
\lesssim 400-500 ~\mbox{Mev}$ propagate in the QGP plasma as
\emph{quasiparticles}, and in particular photons with $k \lesssim
200 ~\mbox{Mev}$ \emph{thermalize} with the plasma on time scales
of the order of $\sim 10-20 ~\mbox{fm}/c$ which is of the order of
the lifetime of the QGP expected at RHIC. }

\item{We predict the photon spectrum for the range of momenta $200
~\mbox{Mev} \lesssim k \lesssim 5~\mbox{Gev}$. The yield is
calculated over a time scale compatible with the expected lifetime
of a QGP at RHIC or LHC $\sim 10~\mbox{fm}/c$.  In the
intermediate region $200 ~\mbox{Mev} \lesssim k \lesssim
2~\mbox{Gev}$ the spectrum is similar to that obtained by the
S-matrix formulation but systematically \emph{larger} by a factor
that ranges between $2-4$. The spectrum flattens at an energy
scale $k \simeq 2.5~\mbox{Gev}$ becoming dramatically
\emph{larger} than the yields obtained previously with the
S-matrix formulation}.

\end{itemize}

The article is organized as follows: in section II we obtain the
real time effective action to lowest order in $\alpha_{em}$ and in
principle to all orders in $\alpha_s$. This effective action
manifestly establishes contact with the stochastic nature of
photon emission from a plasma in thermal equilibrium. In section
III we obtain the photon distribution function in real time in
terms of the solution of a Schwinger-Dyson equation which includes
the photon self-energy to lowest order in $\alpha_{em}$ and in
principle to all orders in $\alpha_s$. This solution provides a
Dyson-like resummation of the naive perturbative expansion. In
section IV we show that the S-matrix result emerges in the strict
perturbative limit. In this section we also make contact with a
kinetic description of photon production which manifestly includes
the dynamics of thermalization. In section V and VI we study the
non-perturbative aspects and obtain the main results of this
article. Section VII summarizes our conclusions and presents
further questions.

\section{The real-time effective action}\label{sec:action}
In refs.\cite{wangphoton,boyanphoton} a manifestly gauge invariant
formulation of the time evolution of an initial density matrix has
been described. We will follow this treatment as it guarantees
that the results are completely gauge invariant.  The gauge
invariant Hamiltonian is given
by\cite{wangphoton,boyanphoton,nuestro}

\begin{equation}\label{totalham}
H =  H_{QCD}[\Psi]+\int
d^3x~\frac{1}{2}~(\vec{E}^2_T+\vec{B}^2)+e\,\int d^3 x ~
 \mathbf{J}\cdot \mathbf{A}_T + H_{coul} \; ,
\end{equation}
\noindent where $H_{QCD}[\Psi]$ is the QCD Hamiltonian in absence
of electromagnetism but in terms of the gauge invariant (under
abelian gauge transformation) quark field(s) $\Psi$ and the
subscript $T$ refers to transverse components. We have extended
the fermion content to $N_f$ flavors and the charge of each flavor
species in units of the electron charge is included in the
 current, namely

 \begin{equation} \mathbf{J} = \sum_{i=1}^{N_f}\frac{e_i}{e} \;
\bar{\Psi}_i \; \vec{\gamma}\; \Psi_i,  \label{currents}
\end{equation}

The instantaneous Coulomb interaction can be traded for a gauge
invariant Lagrange multiplier field which we call $A^0$, (which,
however should \emph{not} be confused with the time component of
the gauge field), leading to the following Lagrangian density

\begin{equation}
\mathcal{L}=
\mathcal{L}_{QCD}+\mathcal{L}_{0,em}-J^0A^0+\mathbf{J}\cdot
\mathbf{A}_T ~~;~~ \mathcal{L}_{0,em}=\frac{1}{2}
\left[(\partial_{\mu} \mathbf{A}_T)^2+ ({\bf \nabla}
A^0)^2\right]~~;~~ {J}^{\mu} =  \sum_{i=1}^{N_f}\frac{e_i}{e} \;
\bar{\Psi}_i \; \gamma^{\mu}\; \Psi_i, \label{4current}
\end{equation}

We will study the time evolution of the  number of physical
transverse photons and the expectation value of the (gauge
invariant) transverse gauge field in the linearized approximation
as an initial value problem.

 In ref.\cite{nuestro} correlated initial states in which quark
 states are dressed by the electromagnetic field were studied in
 detail. One of the important conclusions of that study is that
 such an initial density matrix \emph{cannot} correspond to a QGP
 in LTE under the strong interactions. This is a consequence of
 the fact that correlated quark-photon states necessarily require
 states constructed with the quark-current operator, which does
 not commute with $H_{QCD}$. Furthermore the results of this
 reference  suggest that the details of the initial pre-equilibrium stage
 become manifest in the spectrum for $k \gtrsim 4 ~\mbox{Gev}$.
 The dependence of the high energy part of the spectrum on initial
 conditions has also been studied in\cite{renk} with similar
 conclusions.

Here we study the simpler case of an
 uncorrelated initial density matrix to highlight the non-perturbative aspects,
  with the understanding that
 a precise description of the high energy region will necessarily
 require a firmer knowledge  of the initial state and the
 pre-equilibrium stage.

 Hence we propose the initial density matrix
to be of the form
\begin{equation}
\hat{\rho}(t_i) = \hat{\rho}_{QCD,i} \otimes
\hat{\rho}_{\mathbf{A}_T,i} \label{inidensmtx}
\end{equation}

The initial density matrix  $\hat{\rho}_{QCD,i}$ describes a
quark-gluon plasma in (local) thermodynamic equilibrium at a
temperature $T=1/\beta$, namely

\begin{equation}\label{rhoQCD}
\hat{\rho}_{QCD,i} = e^{-\beta\,H_{QCD}}
\end{equation}

And $\hat{\rho}_{\mathbf{A}_T,i}$ is diagonal in the Fock
representation of free field
 quanta of the transverse gauge field with an initial distribution
 of quanta. It is important to emphasize that any other initial density
matrix that mixes photons and quarks will not commute with
$H_{QCD}$\cite{nuestro}, hence cannot describe a QGP in
equilibrium.

 The vacuum state is represented by
 $\hat{\rho}_{\mathbf{A}_T,i}= |0\rangle\langle0|$ with $|0\rangle$ being
 the vacuum state of the field $\mathbf{A}_T(\vec x)$.

 In the field basis (Schroedinger representation) the matrix
 elements of $\hat{\rho}_{\mathbf{A}_T,i}$ are given by

 \begin{equation}\label{fieldrhoA}
\langle \mathbf{A}_T
|\hat{\rho}_{\mathbf{A}_T,i}|\mathbf{A}^{'}_T\rangle =
\hat{\rho}_{\mathbf{A}_T,i}(\mathbf{A}_T;\mathbf{A}^{'}_T)
 \end{equation}

The time evolution of the initial density matrix is given by

\begin{equation}\label{rhooft}
\hat{\rho}(t_f)= e^{-iH(t_f-t_i)}\hat{\rho}(t_i)e^{iH(t_f-t_i)}
\end{equation}

\noindent where $H$ is the total Hamiltonian given by eq.
(\ref{totalham}).   This particular choice of initial state will
  introduce transient evolution, however the long time behavior
  should be insensitive to this initial transient.

  Furthermore,
  we point out that it is important to study the initial transient
  stage for the following reason. Photons are produced in the thermal bath since the initial density
  matrix does not commute with the total Hamiltonian (non-equilibrium) and they  propagate
  in the medium  as  ``quasiparticles''. By
  studying the initial transient dynamics after  photons and quarks  are coupled we can address the question of the dynamics of
  the \emph{formation} and propagation of the quasiparticle  which will be studied in detail in
  section (\ref{sec:plasmon}).

Our goal is to obtain the real time evolution of the number of
photons produced as well as that of the equation of motion for the
expectation value of the (gauge invariant) transverse gauge field
to lowest order in $\alpha_{em}$ but in principle to \emph{all
orders} in $\alpha_s$. The equation of motion for the expectation
value of the transverse gauge field  will yield real time
information on the formation and propagation of \emph{ transverse
plasmon quasiparticles}(see section \ref{sec:plasmon}).

The strategy is to obtain the real-time effective action for the
transverse gauge fields by integrating out the quark and gluon
degrees of freedom to lowest order in $\alpha_{em}$ but to all
orders in $\alpha_s$. In this manner, the QGP in LTE is treated
effectively as a thermal bath.

The calculation of correlation functions is facilitated by
introducing currents coupled to the different fields. Furthermore
since each time evolution operator in eqn. (\ref{rhooft}) will be
represented as a path integral, we introduce different sources for
the various fields for
 forward and backward time evolution operators, referred generically as
 $J^{+},J^{-}$ respectively.  The
forward and backward time evolution operators in presence of
sources are $U(t_f,t_i;J^+)$, $U^{-1}(t_f,t_i,J^{-})$
respectively.

In order to avoid cluttering of notation let us collectively
denote by $\chi$ the quark and gluon fields which will be
integrated out

In what follows we will ignore the Lagrange multiplier field $A_0$
since the Couloumb interaction will be irrelevant to leading order
in  $\alpha_{em}$.

The non-equilibrium generating functional then is given by
\begin{eqnarray}
{\cal Z}[j^+,j^-]& = &
\mathrm{Tr}U(\infty,t_i;J^+)\hat{\rho}(t_i)U^{-1}(\infty,t_i,J^{-})=
\nonumber \\ &&  \int D\mathbf{A}_{T,i} \int
D\mathbf{A}^{'}_{T,i}\,
\rho_{\mathbf{A}_{T,i}}(\mathbf{A}_{T,i};\mathbf{A}^{'}_{T,i})
\int
 {\cal D}\mathbf{A}^{\pm}_{T}
\int {\cal D} \chi^{\pm}
e^{iS[\mathbf{A}^{\pm}_{T},\chi^{\pm};J^{\pm}_{\mathbf{A}_{T}};J^{\pm}_{\chi}]}
 \label{pathint}\end{eqnarray}
 \noindent where
 \begin{eqnarray}
S[\mathbf{A}^{\pm}_{T},\chi^{\pm};J^{\pm}_{\Phi};J^{\pm}_{\chi}] &
= & \int_{t_i}^{\infty}dt d^3x
\left[\mathcal{L}_{0,em}(\mathbf{A}^+_T)+J^+_{\mathbf{A}}\mathbf{A}^+_T
-\mathcal{L}_{0,em}(\mathbf{A}^-_T)-J^-_{\mathbf{A}}\mathbf{A}^-_T\right]
+\nonumber
\\ &&
\int_{\mathcal{C}}d^4x\left[\mathcal{L}_{QCD}(\chi)+ J\chi
+e\mathbf{J}\cdot\mathbf{A}_T \right] \label{noneqlagradens}
\end{eqnarray}
\noindent and $\mathcal{C}$ describes a contour in the complex
time plane as follows: from $t_i$ to $ +\infty$ (forward) the
fields and sources are $\mathbf{A}^{+}_T$, $\chi^+,J^+_{\chi}$
(where $\chi$ collectively represents both quarks and gluons),
from $+\infty$ back to $t_i$ (backward) the fields and sources are
$\mathbf{A}^{-}_T,\chi^-,J^-_{\chi}$ and from $t_i$ to $
t_i-i\beta$ (Euclidean or Matsubara) the fields and sources are
$\mathbf{A}_T=0; \chi^{\beta},J^{\beta}_{\chi}$. Along the
Euclidean branch the interaction term vanishes since the initial
density matrix for the quark and gluon fields, generically denoted
as $\chi$ is assumed to be that of a QGP in thermal equilibrium.
One can in principle consider a correlated initial state of a QGP
and electromagnetic fluctuations but such state \emph{does not}
describe a QGP in thermal equilibrium and such initial density
matrix would not commute with $H_{QCD}$, contrary to the usual
assumption of an equilibrated QGP\cite{nuestro}.

We seek to obtain the real time effective action for the
transverse gauge field. Therefore we integrate out (trace over)
the quark and gluon fields. After carrying out the trace over the
QCD degrees of freedom, the remaining path integrals over
$\mathbf{A}_T$ with the real time effective action have boundary
conditions on the field $\mathbf{A}_T$  given by

\begin{equation}
\mathbf{A}^+_T(\vec x,t=t_i)= \mathbf{A}_{T,i}(\vec x)  \; \; \; ;
\; \; \; \mathbf{A}^-_T(\vec x,t=t_i)=\mathbf{A}^{'}_{T,i}(\vec x)
\label{condsfi}
\end{equation}

The real time effective action for $\mathbf{A}_T$ is obtained by
treating the quark and gluon fields as a bath performing the path
integral over the quark and gluon degrees of freedom, namely,
tracing over the bath degrees of freedom, leads to the influence
functional\cite{feyver} for $\mathbf{A}^{\pm}_T$.

 The initial
density matrix for the $\mathbf{A}_T$ field will be specified
later as part of the initial value problem.

To lowest order in $\alpha_{em}$ but to \emph{all} orders in
$\alpha_s$ the real time effective action for $\mathbf{A}_T$ is
obtained as follows. Insofar as the path integral over the QCD
degrees of freedom is concerned, $\mathbf{A}_T$ is simply a
background field, hence

\begin{equation}\label{traceQCD}
\int \mathcal{D}\chi^{\pm}
e^{i\int_{\mathcal{C}}\mathcal{L}_{QCD}(\chi^{\pm})+e\mathbf{J}\cdot\mathbf{A}_T}
\equiv \langle e^{i\int_{\mathcal{C}}e\mathbf{J}\cdot\mathbf{A}_T}
\rangle_{QCD}
\end{equation}
\noindent expanding the expectation value in powers of $e$

\begin{equation}
\langle e^{i\int_{\mathcal{C}}e\mathbf{J}\cdot\mathbf{A}_T}
\rangle_{QCD} = 1 + ie \int_{\mathcal{C}} \langle \mathbf{J}
\rangle_{QCD}\cdot\mathbf{A}_T +
\frac{(ie)^2}{2}\int_{\mathcal{C}}\int_{\mathcal{C}}\langle
\mathbf{J}^i
\mathbf{J}^j\rangle_{QCD}\mathbf{A}^i_T\mathbf{A}^j_T+\mathcal{O}(e^3)
= e^{-\frac{e^2}{2}\int_{\mathcal{C}}\int_{\mathcal{C}}\langle
\mathbf{J}^i
\mathbf{J}^j\rangle_{QCD}\mathbf{A}^i_T\mathbf{A}^j_T}+\mathcal{O}(e^3)
\end{equation}
\noindent where we have used $\langle \mathbf{J} \rangle_{QCD}=0$
and the connected current-current correlation function $\langle
\mathbf{J}^i \mathbf{J}^j\rangle_{QCD}$ is in principle to all
orders in $\alpha_s$.

We introduce  the spatial Fourier transform of the quark current
$\mathbf{J}$ as
\begin{equation}
\mathbf{j}(\vec k;t) = \frac{1}{\sqrt{\Omega}} \int d^3x e^{i \vec
k \cdot \vec x}
 \mathbf{J}(\vec x,t)
\label{spatialFT}
\end{equation}
with $\Omega$ the quantization volume, in terms of which we obtain
following the correlation functions
\begin{eqnarray}
e^2\langle \mathbf{j}_{l}(\vec k;t) \mathbf{j}_m(-\vec k;
t')\rangle & = & e^2\langle \mathbf{j}^-_{l}(\vec k;t)
\mathbf{j}^+_{m}(-\vec k;t')\rangle =
{\Sigma}^>_{lm}(\vec{k};t-t')
=  {\Sigma}^{-+}_{lm}(\vec k;t,t')\label{pigreat} \\
e^2\langle \mathbf{j}_m(-\vec k; t') \mathbf{j}_{l}(\vec
k;t)\rangle & = & e^2\langle \mathbf{j}^+_{l}(\vec k;t)
\mathbf{j}^-_{m}(-\vec k;t')\rangle ={\Sigma}^<_{lm}(\vec{k};t-t')
=
{\Sigma}^{+-}_{lm}(\vec k;t,t')= {\Sigma}^{-+}_{ml}(\vec{k};t',t)\label{pilesser} \\
e^2\langle T~ \mathbf{j}_{l}(\vec k;t) \mathbf{j}_m(-\vec k;
t')\rangle & = & {\Sigma}^>_{lm}(\vec{k};t-t')\Theta(t-t')+
{\Sigma}^<_{lm}(\vec k;t-t')\Theta(t'-t)= {\Sigma}^{++}_{lm}(\vec{k};t,t') \label{timeordered} \\
e^2\langle \tilde{T}~ \mathbf{j}_{l}(\vec k;t) \mathbf{j}_m(-\vec
k; t')\rangle & = & {\Sigma}^>_{lm}(\vec{k};t-t')\Theta(t'-t)+
{\Sigma}^<_{lm}(\vec{k};t-t')\Theta(t-t') =
{\Sigma}^{--}_{lm}(\vec{k};t,t')\label{antitimeordered}
\end{eqnarray}
where  the expectation values  are in the equilibrium thermal
density matrix of the QGP, which results in  translational and
rotational invariant  correlation functions that are only
functions of the time differences and the superscripts $\pm$ refer
to the forward ($+$) or backward $(-)$ time branches. The symbols
$T$ and $\tilde{T}$ refer to time and antitime ordering
respectively.  These correlation functions are not independent,
they obey

\begin{equation}\label{const}
{\Sigma}^{++}_{lm}(\vec{k};t,t') +
\Sigma^{--}_{lm}(\vec{k};t,t')-\Sigma^{-+}_{lm}(\vec{k};t,t')-\Sigma^{+-}_{lm}(\vec{k};t,t')=0
\end{equation}

It is convenient at this stage to separate the transverse and
longitudinal components of the polarization tensor
$\Sigma_{lm}(\vec{k};t-t')$ as follows

\begin{eqnarray}
\Sigma_{lm}(\vec{k};t-t')= &&
\mathcal{P}_{lm}(\hat{\mathbf{k}})\Sigma_T(\vec k;t-t')+
\hat{\mathbf{k}}_l\hat{\mathbf{k}}_m
\Sigma_L(\vec k;t-t')\label{polasplit}\\
\mathcal{P}_{lm}(\hat{\mathbf{k}})=&&
\delta_{lm}-\hat{\mathbf{k}}_l\hat{\mathbf{k}}_m \label{transproj}
\end{eqnarray}

The  non-equilibrium real time effective action  in terms of the
spatial Fourier transforms of the fields and correlation functions
to lowest order in $\alpha_{em}$ but to \emph{all orders } in
$\alpha_s$ is therefore  given by

\begin{eqnarray}\label{influfunc}
iS_{eff}[\mathbf{A}^+_T,\mathbf{A}^-_T] & = & \sum_{\vec k}\left\{
\frac{i}{2} \int dt \left[\dot{\mathbf{A}}^+_{\vec k,T}(t)\cdot
\dot{\mathbf{A}}^+_{-\vec k,T}(t)- k^2\mathbf{A}^+_{\vec
k,T}(t)\cdot\mathbf{A}^+_{-\vec k,T}(t) \right. \right.\nonumber\\
 &  & \left. \left. -\dot{\mathbf{A}}^-_{\vec k,T}(t)\dot{\mathbf{A}}^-_{-\vec k,T}(t)+
 k^2\mathbf{A}^-_{\vec k,T}(t)\mathbf{A}^-_{-\vec k,T}(t) \right]
  \right. \nonumber \\
&   & \left. - \frac{1}{2} \int dt \int dt' \left[
\mathbf{A}^+_{\vec k,T}(t)\cdot \mathbf{A}^+_{-\vec
k,T}(t'){\Sigma}^{++}_{T}(\vec k;t-t')+ \mathbf{A}^-_{\vec
k,T}(t)\cdot\mathbf{A}^-_{-\vec k,T}(t')
\Sigma^{--}_T(\vec k;t-t') \right. \right. \nonumber \\
&& \left. \left. -\mathbf{A}^+_{\vec k,T}(t)\cdot
\mathbf{A}^-_{-\vec k,T}(t'){\Sigma}^{+-}_{T}(\vec k;t-t')-
\mathbf{A}^-_{\vec k,T}(t)\cdot \mathbf{A}^+_{-\vec
k,T}(t'){\Sigma }^{-+}_T(\vec k;t-t')\right] \right\}
\end{eqnarray}

  As it will become clear below,
it is more convenient to introduce the following Wigner center of
mass and relative variables for the transverse gauge field
\begin{equation}
\vec{\mathcal{A}}(\vec x,t)  =  \frac{1}{2} \left(
\mathbf{A}^+_T(\vec x,t) + \mathbf{A}^-_T(\vec x,t) \right) \; \;
; \; \; \mathbf{a}(\vec x,t) = \mathbf{A}^+_T(\vec x,t) -
\mathbf{A}^-_T(\vec x,t)  \label{wigvars}
\end{equation}
and the Wigner transform of the initial density matrix for the
transverse gauge  field
\begin{equation}
{\cal W}(\vec{\mathcal{A}}_i ; \vec{\mathcal{E}}_i) = \int D
\vec{a}_i \; e^{-i\int d^3x \vec{\mathcal{E}}_i(\vec x)\cdot
\vec{a}_i(\vec x)}
\rho(\vec{\mathcal{A}}_i+\frac{\vec{a}_i}{2};\vec{\mathcal{A}}_i-\frac{\vec{a}_i}{2})~~;~~
\rho(\vec{\mathcal{A}}_i+\frac{\vec{a}_i}{2};\vec{\mathcal{A}}_i-\frac{\vec{a}_i}{2})=
\int D \vec{\mathcal{E}}_i \; e^{i\int d^3x
\vec{\mathcal{E}}_i(\vec x)\cdot \vec{a}_i(\vec x)}{\cal
W}(\vec{\mathcal{A}}_i ; \vec{\mathcal{E}}_i) \label{wignerrho}
\end{equation}
The initial conditions on the $\mathbf{A}_T$ path integral given
by (\ref{condsfi}) translate into the following initial conditions
on the center of mass and relative variables
\begin{equation}
\vec{\mathcal{A}}(\vec x,t=0)= \vec{\mathcal{A}}(\vec x)\; \; ; \;
\; \mathbf{a}(\vec x,t=0)=\mathbf{a}_i \label{bcwig}
\end{equation}

The center of mass variable plays an important role: its
expectation value is the mean field, since the expectation values
of $\vec{A}^\pm_T(\vec x,t)$ coincide,

\begin{equation}
\langle \vec{A}^+_T(\vec x,t) \rangle = Tr \vec{A}_T(\vec x,t)
~\rho = Tr \rho ~\vec{A}_T(\vec x,t) = \langle \vec{A}^-_T(\vec
x,t) \rangle
\end{equation}

In terms of the spatial Fourier transforms of the center of mass
and relative  variables (\ref{wigvars}) introduced above,
integrating  by parts and accounting for the boundary conditions
(\ref{bcwig})  the non-equilibrium effective action
(\ref{influfunc}) becomes:
\begin{eqnarray}
iS_{eff}[\vec{\mathcal{A}},\mathbf{a}] & = & \int dt \sum_{\vec k}
\left\{-i~\vec{a}_{-\vec k}(t)\cdot
\left( \ddot{\vec{\mathcal{A}}}_{\vec k}(t)+k^2\vec{\mathcal{A}}_{\vec k}(t) \right)\right\} \nonumber \\
                 & - & \int dt \int dt' \left\{\frac{1}{2}\vec{a}_{-\vec k}(t)\cdot \vec{a}_{\vec k}(t'){\cal K}_T(k;t-t') + \vec{a}_{-\vec k}(t)
 \cdot \vec{\mathcal{A}}_{\vec k}(t')~ i\Sigma^R_T(k;t-t')\right \} \nonumber \\
& + & \int d^3x \vec{a}_i(\vec x) \cdot
\dot{\vec{\mathcal{A}}}(\vec x,t=0) \label{efflanwig}
\end{eqnarray}
where the last term arises after the integration by parts in time,
using the boundary condition (\ref{bcwig}). There is no
contribution from the  $t \rightarrow \infty$ limit since the
fields $\vec{\mathcal{A}}_{\vec k}(t)$ at non-zero temperature
will vanish at asymptotically long time. The kernels in the above
effective Lagrangian are given by
\begin{eqnarray}
\mathcal{K}_T(k;t-t') & = &
\frac{1}{2} \left[{\Sigma}_T^>(k;t-t')+{\Sigma}_T^<(k;t-t') \right] \label{kernelkappa} \\
i\Sigma^{R}_T(k;t-t') & = &
\left[{\Sigma}_T^>(k;t-t')-{\Sigma}_T^<(k;t-t')
\right]\Theta(t-t') \equiv i\Sigma_T(k;t-t')\Theta(t-t')
\label{kernelsigma}
\end{eqnarray}
\noindent where we have used the fact that the polarization is a
function of the modulus of the wavevector by rotational
invariance.  The photon polarization is computed to lowest order
in $\alpha_{em}$ and in principle to all orders in $\alpha_s$.

 The  quadratic part in $e^{iS_{eff}[\vec{\mathcal{A}},\mathbf{a}]
 }$ in
the relative variable $\vec{a}$ can be written in terms of a
stochastic noise variable $\xi$ as
\begin{eqnarray}
\exp\Bigg\{-\frac{1}{2} \int dt \int dt' \vec{a}_{-\vec k}(t)\cdot
\vec{a}_{\vec k}(t')\mathcal{K}_T(k;t-t')\Bigg\} = \int {\cal
D}\vec{\xi} \exp\left\{-\frac{1}{2} \int dt \int dt' ~~
{\xi}_{\vec k,l}(t) {\mathcal K}^{-1}_T(t-t') {\xi}_{-\vec
k,l}(t')
 \right. \nonumber \\
\left. +i \int dt ~~ {\xi}_{-\vec k,l}(t)\cdot {a}_{\vec
k,l}(t)\right\} \label{noisefunc}
\end{eqnarray}

The non-equilibrium generating functional can now be written  as
\begin{eqnarray}
{\cal Z}  & = &   \int D \vec{\mathcal{A}}_i \int D
\vec{\mathcal{E}}_i \int {\mathcal D}\vec{\mathcal{A}} {\mathcal
D}\vec{a} {\mathcal D}\vec{\xi} ~~ {\cal
W}(\vec{\mathcal{A}}_i;\vec{\mathcal{E}}_i) D~\vec{a}_i
 e^{i \int d^3x \vec{a}_i(\vec x) \cdot \left(\vec{\mathcal{E}}_i (\vec x)-\dot{\vec{\mathcal{A}}}(\vec x,t=0)\right)}
 {\cal P}[\xi] \label{genefunc} \\
&  & \exp\left\{-i \int dt \sum_{\vec k}~ a_{-\vec k,l}(t) \left[
\ddot{\mathcal{A}}_{\vec k,l}(t)+k^2\mathcal{A}_{\vec k,l}(t)+\int dt' ~~
 \Sigma^{R}_T(k;t-t')\mathcal{A}_{\vec k,l}(t')-\xi_{\vec k,l}(t) \right] \right\} \nonumber
 \end{eqnarray}

\noindent where  the noise probability distribution function is
given by
 \begin{equation}
{\cal P}[\vec{\xi}]  =  \exp\left\{-\frac{1}{2} \int dt \int dt'
\sum_{\vec k} ~~ \xi_{\vec k,l}(t) {\cal K}^{-1}_T(k;t-t')
\xi_{-\vec k,l}(t') \right\} \label{probaxi}
\end{equation}

The functional integral over $\vec{a}_i$ can now be done,
resulting in a functional delta function, that fixes the initial
condition $\dot{\mathcal{A}}_T(\vec x,t=0) =
\mathcal{A}_{T,i}(\vec x)$.

Finally the path integral over the relative variable $\vec{a}$
can be performed leading to a functional delta function and the
final form of the generating functional is given by
\begin{eqnarray}
{\cal Z}   =   \int D \vec{\mathcal{A}}_i D \vec{\mathcal{E}}_i
~~{\cal W}(\vec{\mathcal{A}}_i;\vec{\mathcal{E}}_i)
 {\cal D} \vec{\mathcal{A}} {\cal D}\xi ~~ {\cal P}[\xi]~\Pi_{\vec
 k,l}
 \delta\left[
\ddot{\mathcal{A}}_{\vec k,l}(t)+k^2\mathcal{A}_{\vec
k,l}(t)+\int^t_0 dt' ~~ \Sigma_T(k;t-t') \mathcal{A}_{\vec
k,l}(t')-\xi_{\vec k,l}(t) \right]  \label{deltaprob}
\end{eqnarray}
with the initial conditions on the path integral on $\Psi$ given
by
\begin{equation}
\vec{\mathcal{A}}(\vec x,t=0) = \vec{\mathcal{A}}_{i}(\vec x) \;
\; ; \; \; \dot{\vec{\mathcal{A}}}(\vec x,t=0)=
\vec{\mathcal{E}}_i(\vec x) \label{bcfin}
\end{equation}
\noindent and we have used the definition of
$\Sigma^{R}_T(k;t-t')$ in terms of $\Sigma_T(k;t-t')$ given in
equation (\ref{kernelsigma}).

The meaning of the above generating functional is the following:
in order to obtain the correlation functions of the center of mass
Wigner variable $\vec{\mathcal{A}}$ we must first find the
solution of the  {\em classical stochastic} non-local Langevin
equation of motion
\begin{eqnarray}
&& \ddot{\mathcal{A}}_{\vec k,l}(t)+k^2\mathcal{A}_{\vec
k,l}(t)+\int^t_0 dt' ~~ \Sigma_T(k;t-t') \mathcal{A}_{\vec
k,l}(t')= \xi_{\vec k,l}(t)  \nonumber \\
&& \vec{\mathcal{A}}(\vec x,t=0) = \vec{\mathcal{A}}_{i}(\vec x)
\; \; ; \; \; \dot{\vec{\mathcal{A}}}(\vec x,t=0)=
\vec{\mathcal{E}}_i(\vec x)  \label{langevin}
\end{eqnarray}
\noindent for arbitrary noise term $\xi$ and then average the
products of $\vec{\mathcal{A}}[\xi]$ over the stochastic noise
with the probability distribution $ {\cal P}[\xi]$ given by
(\ref{genefunc}), and finally average over the initial
configurations $\vec{\mathcal{A}}_{i}(\vec x) \;;\;
\vec{\mathcal{E}}_i(\vec x)$ weighted by the Wigner function
${\cal W}(\vec{\mathcal{A}}_i;\vec{\mathcal{E}}_i)$, which plays
the role of an initial semiclassical phase space distribution
function.

 Calling
the solution of (\ref{langevin})
$\vec{\mathcal{A}}_{\vec{k},l}(t;\vec{\xi};\vec{\mathcal{A}}_i;\vec{\mathcal{E}}_i)$,
 the two point correlation function, for example, is given by
\begin{eqnarray}
\langle \vec{\mathcal{A}}_{-\vec{k},l}(t)
\vec{\mathcal{A}}_{\vec{k},l}(t') \rangle = &&
\mathcal{Z}^{-1}\int {\cal D}[\xi] {\cal P}[\xi] \int D
\vec{\mathcal{A}}_i \int D\vec{\mathcal{E}}_i~~{\cal
W}(\vec{\mathcal{A}}_i;\vec{\mathcal{E}}_i)~
\vec{\mathcal{A}}_{-\vec{k},l}(t;\vec{\xi};\vec{\mathcal{A}}_i;\vec{\mathcal{E}}_i)\,
\vec{\mathcal{A}}_{\vec{k},l}(t';\vec{\xi};\vec{\mathcal{A}}_i;\vec{\mathcal{E}}_i)\nonumber
\\ && \times \delta\left[ \ddot{\mathcal{A}}_{\vec
k,l}(t)+k^2\mathcal{A}_{\vec k,l}(t)+\int^t_0 dt' ~~
\Sigma_T(k;t-t') \mathcal{A}_{\vec k,l}(t')-\xi_{\vec k,l}(t)
\right] \label{expecvalcm}
\end{eqnarray}

We note that in computing the averages and using the functional
delta function to constrain the configurations of
$\vec{\mathcal{A}}$ to the solutions of the Langevin equation,
there is the Jacobian of the operator

\begin{equation}
\delta(t-t')\left[\frac{d^2}{dt^2} +k^2\right]+
\Sigma^{R}_{T}(k;t-t')
\end{equation}
\noindent which, however, is independent of the field and cancels
between numerator and denominator in the averages.

This formulation establishes the connection with a
\emph{stochastic} problem and is similar to the
Martin-Siggia-Rose\cite{MSR} path integral formulation for
stochastic phenomena. There are two different averages:

\begin{itemize}
\item{ The average over the  stochastic noise term, which up to
this order is Gaussian. We denote the averages over the noise with
the probability distribution function $P[\xi]$ given by eqn.
(\ref{probaxi}) as

\begin{equation}\label{stocha}
\langle \langle \mathcal{O}[\xi] \rangle \rangle \equiv \frac{\int
\mathcal{D}\xi P[\xi] \mathcal{O}[\xi]}{\int \mathcal{D}\xi
P[\xi]}. \end{equation}

Since the noise probability distribution function is Gaussian the
only necessary correlation functions for the noise are given by

\begin{equation}
\langle \langle \xi_{\vec{k},l}(t)\rangle \rangle =0 \; , \;
\langle \langle \xi_{\vec{k},l}(t)\xi_{\vec{k}',j}(t')\rangle
\rangle = \mathcal{P}_{lj}(\hat{\mathbf{k}})~{\mathcal
K}_T(k;t-t')\delta^{3}(\vec{k}+\vec{k}') \label{noisecorrel}
\end{equation}
\noindent and the higher order correlation functions are obtained
from Wick's theorem.
 }

\item{Average over the initial conditions with the Wigner
distribution function ${\cal
W}(\vec{\mathcal{A}}_i;\vec{\mathcal{E}}_i)$ which we denote as

\begin{equation}
\overline{\mathcal{O}[\vec{\mathcal{A}}_i;\vec{\mathcal{E}}_i]}
\equiv  \frac{\int D \vec{\mathcal{A}}_i \int D\vec{\mathcal{E}}_i
~~{\cal W}(\vec{\mathcal{A}}_i;\vec{\mathcal{E}}_i)
\mathcal{O}[\vec{\mathcal{A}}_i;\vec{\mathcal{E}}_i]}{\int D
\vec{\mathcal{A}}_i \int D\vec{\mathcal{E}}_i ~~{\cal
W}(\vec{\mathcal{A}}_i;\vec{\mathcal{E}}_i) } \label{iniaverage}
\end{equation}
The initial Wigner distribution function requires precise
knowledge of the pre-equilibrium stage. It allows to include
initial correlations of quarks and photons, namely ``entangled
states'' but as discussed in detail in ref.\cite{nuestro} such
initial density matrix will not commute with the $H_{QCD}$
Hamiltonian and will not describe a state in LTE under the strong
interactions. All of the theoretical uncertainties about the
initial state prior to equilibration are encoded in the initial
density matrix or alternatively in the initial Wigner distribution
function ${\cal W}(\vec{\mathcal{A}}_i;\vec{\mathcal{E}}_i)$. In
ref.\cite{nuestro} a particular state with initial correlations
reflecting a pre-equilibrium stage was modelled. While we can
consider such modelled initial state, in this article we focus on
extracting non-perturbative aspects in the simplest and cleanest
scenario, that of an initially uncorrelated Gaussian state for
photons. The numerical study performed in ref.\cite{nuestro}
revealed that initial state preparation on time scales of
$\mathcal{O}(1 ~\mbox{fm}/c)$ modifies the photon spectrum for
hard momenta $k \gtrsim 4~\mbox{Gev}$. This is roughly the scale
of momenta at which the assumption of LTE breaks down in any case
because the photon probes scales of the order of the mean free
path for quark-gluon collisions. Thus studying an uncorrelated
initial state, which is compatible with all previous  S-matrix
calculations is phenomenologically relevant for photon momenta $k
\leq 4-5 ~\mbox{Gev}$.

Therefore in what follows we will consider a Gaussian initial
Wigner distribution function with the following averages:

\begin{eqnarray}
&&\overline{{\mathcal{A}}^a_{i,\vec k}{\mathcal{A}}^b_{i,-\vec k}}
=
\frac{\mathcal{P}^{ab}(\hat{\mathbf{k}})}{2k}\left[1+2\mathcal{N}_k\right]+
\overline{{\mathcal{A}}^a_{i,\vec
k}}~~~\overline{{\mathcal{A}}^b_{i,-\vec k}} ~;~\label{psi2}\\
&&\overline{{\mathcal{E}}^a_{i,\vec k}{\mathcal{E}}^b_{i,-\vec k}}
=\mathcal{P}^{ab}(\hat{\mathbf{k}})
\frac{k}{2}\left[1+2\mathcal{N}_k\right]+\overline{{\mathcal{E}}^a_{i,\vec
k}}~~~\overline{{\mathcal{E}}^b_{i,-\vec k}} ~;~\label{pi2}\\
&&\overline{{\mathcal{A}}^a_{i,\vec k}{\mathcal{E}}^b_{i,-\vec
k}+{\mathcal{E}}^b_{i,-\vec k}{\mathcal{A}}^a_{i,\vec
k}}=\overline{{\mathcal{A}}^a_{i,\vec
k}}~~\overline{{\mathcal{E}}^b_{i,-\vec k}}+\mathrm{c.c}
\end{eqnarray}

\noindent where $\mathcal{N}_k$ is the initial distribution of
photons. The indices $a,b$ refer to vector components, while the
label $i$ refers to the initial conditions (\ref{bcfin}). }

\end{itemize}

Thus   averages in the time evolved full density matrix are
therefore given by

\begin{equation}
\langle \cdots \rangle = \overline{\langle \langle \cdots\rangle
\rangle}~~. \label{totave}
\end{equation}

The noise field emerged as an auxiliary variable to represent the
quadratic contribution from the relative variable in eqn.
(\ref{efflanwig}). While the full microscopic dynamics is
completely Hamiltonian and therefore deterministic, integrating
out the QGP degrees of freedom leads to a reduced density matrix
 and the ensuing coarse grained dynamics for  the photon field,
 just as in the microscopic treatment of Brownian
 motion\cite{feyver}.

\subsection{Relation between $\Sigma$ and $\mathcal{K}$ : Fluctuation and Dissipation}

 From the expression (\ref{kernelsigma}) for the photon polarization  which is determined by
averages in the  equilibrium density matrix of QCD we now obtain a
dispersive representation for the kernels $\mathcal{K}_T(k;t-t')$
and $\Sigma^{R}_T(k;t-t')$.
 This is achieved by explicitly writing the expectation value in
terms of energy eigenstates of the bath (quark and gluon fields)
introducing the identity in this basis and using the time
evolution of the Heisenberg field operators to obtain
\begin{eqnarray}
{\Sigma}_T^>({k};t-t') & = &  \int^{\infty}_{-\infty} d\omega ~ \sigma^>_{T}( k;\omega)~e^{i\omega(t-t')}  \label{specrepgreat} \\
{\Sigma}_T^<({k};t-t') & = &  \int^{\infty}_{-\infty} d\omega ~
\sigma^<_{T}( k;\omega)~e^{i\omega(t-t')} \label{specrepless}
\end{eqnarray}

\noindent with the spectral functions

\begin{eqnarray}
\sigma^>_{T}(k;\omega) & = & e^2\frac{\mathcal{P}_{ij}(\hat{\bf
k})}{2Z_{QCD}} \sum_{m,n}e^{-\beta E_n}
\langle n| \mathbf{j}_i(\vec{k},0) |m \rangle \langle m| \mathbf{j}_j(-\vec{k},0) |n \rangle \, \delta(\omega-(E_n-E_m)) \label{siggreat} \\
\sigma^<_{T}( k;\omega) & = & e^2 \frac{\mathcal{P}_{ij}(\hat{\bf
k})}{2Z_{QCD}} \sum_{m,n} e^{-\beta E_n}
 \langle n| \mathbf{j}_j(-\vec{k},0) |m \rangle \langle m| \mathbf{j}_i(\vec{k},0) |n
 \rangle \, \delta(\omega-(E_m-E_n))
 \label{sigless}
\end{eqnarray}

\noindent where $Z_{QCD}$ is the QCD equilibrium partition
function, the states $|n>$ are exact eigenstates of $H_{QCD}$ and
we have used rotational invariance. Upon relabelling $m
\leftrightarrow n$ in the sum in eq.  (\ref{sigless}) we find the
KMS relation\cite{kapustabook,lebellac}

\begin{equation}
\sigma^<_{ T}(k;\omega)  = \sigma^>_{ T}(k;-\omega) = e^{\beta
\omega} \sigma^>_{T}(k;\omega) \label{KMS}
\end{equation}

\noindent where we have used parity and rotational invariance in
the second line above to assume that the spectral functions only
depend of the absolute value of the momentum.

Using the spectral representation of the $\Theta(t-t')$ we find
the following representation for the retarded photon polarization
given by eqn. (\ref{kernelsigma})

\begin{equation}
\Sigma^R_T(k;t-t')= \int_{-\infty}^{\infty}\frac{dk_0}{2\pi}
e^{ik_0(t-t')} \tilde{\Sigma}^R_T(k,k_0) \label{sigreta}
\end{equation}

\noindent with

\begin{equation}\label{sigofomega}
\tilde{\Sigma}^R_T(k,k_0)=\int_{-\infty}^{\infty}d\omega
\frac{[\sigma^>_k(\omega)-\sigma^<_k(\omega)]}{\omega-k_0+i\epsilon}
\end{equation}

Using the condition (\ref{KMS}) the above spectral representation
can be written in a more useful manner as

\begin{eqnarray}
\tilde{\Sigma}^R_T(k,k_0) & = &
-\frac{1}{\pi}\int_{-\infty}^{\infty}d\omega
\frac{\mathrm{Im}\tilde{\Sigma}^R_T(k,\omega)}{\omega-k_0+i\epsilon}\label{specsigret}\\
\mathrm{Im}\tilde{\Sigma}^R_T(k,\omega) & = & \pi
\sigma^>_T(k;\omega)\left[e^{\beta \omega}-1\right]
\label{imagpart}
\end{eqnarray}

\noindent clearly $\mathrm{Im}\tilde{\Sigma}^R_T(k;\omega>0)>0$.
Eq. (\ref{KMS}) entails that the imaginary part of the retarded
photon polarization is an odd function of frequency, namely

\begin{equation}
\mathrm{Im}\tilde{\Sigma}^R_T(k,\omega) = -
\mathrm{Im}\tilde{\Sigma}^R_T(k,-\omega) \label{odd}
\end{equation}

\noindent which is manifest in eq. (\ref{imagpart}).

The relation (\ref{imagpart}) leads to the  following  results
which will prove to be useful later

\begin{eqnarray}
\sigma^>_T(k;\omega)& = &
\frac{1}{\pi}\mathrm{Im}\tilde{\Sigma}^R_T(k,\omega)\,n(\omega)
\label{sigplu}\\
\sigma^<_T(k;\omega)& = &
\frac{1}{\pi}\mathrm{Im}\tilde{\Sigma}^R_T(k;\omega)\,\left[1+n(\omega)\right]\label{sigmin}
\end{eqnarray}

Similarly from the eqs.  (\ref{kernelkappa}) and
(\ref{specrepgreat}), (\ref{specrepless}) and the condition
(\ref{KMS}) we find

\begin{eqnarray}
\mathcal{K}_T(k;t-t') & =
&\int_{-\infty}^{\infty}\frac{dk_0}{2\pi}
e^{ik_0(t-t')} \tilde{\mathcal{K}}_T(k;k_0) \label{noisefou} \\
\tilde{\mathcal{K}}_T(k;k_0)& = & \pi
\sigma^>_T(k;k_0)\left[e^{\beta k_0}+1\right] \label{noisekernel}
\end{eqnarray}

\noindent whereupon using the condition (\ref{KMS}) leads to the
generalized form of the fluctuation-dissipation relation

\begin{equation}
\tilde{\mathcal{K}}_T(k;k_0)=\mathrm{Im}\tilde{\Sigma}^R_T(k;k_0)\coth\left[\frac{\beta
k_0}{2}\right]\label{flucdiss}
\end{equation}

Thus we see that
$\mathrm{Im}\tilde{\Sigma}^R_T(k;k_0)\,;\,\tilde{\mathcal{K}}_T(k;k_0)$
are odd and even functions of the frequency respectively.

For further analysis below we will also need the following
representation for $\Sigma_T(k;t-t')$ introduced in eq.
(\ref{kernelsigma})

\begin{equation}
\Sigma_T(k;t-t') = -i \int_{-\infty}^{\infty} e^{i\omega(t-t')}
\left[\sigma^>_T(k;\omega)-\sigma^<_T(k;\omega)\right]d\omega =
\frac{i}{\pi}\int_{-\infty}^{\infty}
e^{i\omega(t-t')}\mathrm{Im}\tilde{\Sigma}^R_T(k;\omega) d\omega
\label{sig}
\end{equation}

\noindent whose Laplace transform is given by

\begin{equation}
\tilde{\Sigma}_T(k;s)\equiv \int^{\infty}_0 dt
e^{-st}\Sigma_T(k;t)= -\frac{1}{\pi} \int^{\infty}_{-\infty}
\frac{\mathrm{Im}\tilde{\Sigma}^R_T(k;\omega)}{\omega+is}d\omega
\label{laplasig}
\end{equation}

This spectral representation, combined with (\ref{specsigret})
lead to the relation

\begin{equation}
\tilde{\Sigma}^R_T(k;\omega)=\tilde{\Sigma}_T(k;s=i\omega+0)\label{analyt}
\end{equation}

\section{The photon distribution function in real time}

The solution of the Langevin equation (\ref{langevin}) can be
found by Laplace transform. Defining the Laplace transforms

\begin{equation}\label{laplapsi}
\tilde{\mathcal{A}}_{\vec k,l}( s)  \equiv \int^{\infty}_0 dt
e^{-st}\mathcal{A}_{\vec k,l}(t)~~;~~ \tilde{\xi}_{\vec k,l}(s)
\equiv \int^{\infty}_0 dt e^{-st}\xi_{\vec k,l}(t)
\end{equation}

\noindent along with the Laplace transform of the photon
polarization given by eqn. (\ref{laplasig}) we find the solution

\begin{equation}\label{solution}
\tilde{\mathcal{A}}_{\vec k,l}( s)= \frac{\mathcal{E}^i_{\vec
k,l}+s \mathcal{A}^i_{\vec k,l}+\tilde{\xi}_{\vec
k,l}(s)}{s^2+k^2+\tilde{\Sigma}_T(k;s)}
\end{equation}

\noindent where

\begin{equation}
\mathcal{A}^i_{\vec k,l}= \frac{1}{\sqrt{\Omega}}\int d^3x
e^{i\vec k \cdot \vec x} \mathcal{A}^i_{l}(\vec x) ~~;~~
\mathcal{E}^i_{\vec k,l}=\frac{1}{\sqrt{\Omega}}\int d^3x e^{i\vec
k \cdot \vec x} \mathcal{E}^i_{l}(\vec x)\end{equation}

\noindent and the superscript $i$ refers to the initial conditions
(\ref{bcfin}). The solution in real time can be written in a more
compact manner as follows. Introduce the fundamental solution
$f_k(t)$ of the equation of motion

\begin{equation}
\ddot{f}_{k}(t)+k^2 \,f_{ k}(t)+\int_0^t dt' ~ \Sigma_{T}(k;t-t')
f_{ k}(t')=0 \label{functionf}
\end{equation}
\noindent obeying the initial conditions\footnote{The lower limit
in the integral $t=0$ simply reflects the choice of the initial
time. If an arbitrary initial time is chosen $t_0$, the lower
limit becomes $t_0$, since the $\Sigma_T(k;t-t')$ is manifestly
time translational invariant, the solution of the equation of
motion is a function of $t-t_0$. }

\begin{equation}
 f_{k}(t=0)= 0; ~~ \dot{f}_{ k}(t=0)=1
 \end{equation}

Its Laplace transform is given by

\begin{equation}\label{laplaf}
\tilde{f}_k(s) = \frac{1}{s^2+k^2+\tilde{\Sigma}_T(k;s)}
\end{equation}

\noindent which is recognized as the Laplace transform of the full
transverse photon propagator.

The fundamental solution  $f_k(t)$ is found by the inverse Laplace
transform

\begin{equation}\label{bromw}
f_k(t) = \int_{C}\frac{ds}{2\pi i} \frac{e^{st}
}{s^2+k^2+\tilde{\Sigma}_T(k;s)}
\end{equation}
\noindent where $C$ stands for the Bromwich contour, parallel to
the imaginary axis in the complex $s$ plane to the right of all
the singularities of $\tilde{f}(s)$ and along the semicircle at
infinity for $\mathrm{Re}\,s < 0$. The singularities of
$\tilde{f}(s)$ in the physical sheet are isolated single particle
poles  and multiparticle cuts along the imaginary axis. Thus the
contour runs parallel to the imaginary axis with a small positive
real part with $s=i\omega+\epsilon\,;\,-\infty \leq \omega \leq
\infty$ and wraps around returning parallel to the imaginary axis
with $s=i\omega-\epsilon\,;\,\infty>\omega>-\infty$, with
$\epsilon = 0^+$. From the spectral representations
(\ref{imagpart},\ref{laplasig})) one finds that
$\tilde{\Sigma}_T(k,s=i\omega\pm\epsilon)=\mathrm{Re}\tilde{\Sigma}^R_T(k,\omega)\pm
\mathrm{Im}\tilde{\Sigma}^R_T(k,\omega)$ and using that
$\mathrm{Im}{\Sigma}^R_T(k,\omega)=-\mathrm{Im}\tilde{\Sigma}^R_T(k,-\omega)$
we find

\begin{equation}\label{foft}
f_k(t)= \int_{-\infty}^{\infty} \frac{d\omega}{\pi}
\frac{\sin(\omega
t)~\left[\mathrm{Im}\tilde{\Sigma}^R_T(k;\omega)+2\omega\epsilon\right]}{\left[\omega^2-k^2-\mathrm{Re}\tilde{\Sigma}^R_T(k;\omega)
\right]^2+\left[\mathrm{Im}\tilde{\Sigma}^R_T(k;\omega)+2\omega\epsilon\right]^2}
\end{equation}

We have kept the infinitesimal $2\omega\epsilon\,;\,\epsilon
\rightarrow 0^+$ to highlight the possibility of isolated
quasiparticle poles in the case of vanishing imaginary part of the
polarization.

The initial condition $\dot{f}_k(t=0)=1$ results in the following
sum rule
\begin{equation}
\int_{-\infty}^{\infty} \frac{d\omega}{\pi} ~\frac{\omega
\left[\mathrm{Im}\tilde{\Sigma}^R_T(k;\omega)+2\omega\epsilon\right]}{\left[\omega^2-\omega^2_k-\mathrm{Re}\tilde{\Sigma}^R_T(k;\omega)
\right]^2+\left[\mathrm{Im}\tilde{\Sigma}^R_T(k;\omega)+2\omega\epsilon\right]^2}
=1 \label{sumrule}
\end{equation}

In terms of the  fundamental solution $f_k(t)$ given above,  the
solution of the Langevin equation (\ref{langevin}) in real time is
given by

\begin{equation}
\mathcal{A}_{\vec k,l}(t;\mathcal{A}_i;\mathcal{E}_i;\xi) =
\mathcal{A}^i_{\vec k,l}~ \dot{f}_k(t) +
 \mathcal{E}^i_{\vec
k,l}~ f_k(t)+ \int^t_0 f_k(t-t')~\xi_{\vec k,l}(t') dt'
\label{inhosolution}
\end{equation}

\noindent where the superscript $i$ refers to the initial
conditions (\ref{bcfin}).

This solution in real time represents the resummation of the Dyson
series.

\subsection{Number operator:}
We \emph{define} the photon number operator   to be given by
\begin{equation}
\sum_{\lambda}\hat{N}_{k,\lambda}(t) = \frac{1}{2k\,Z} \left \{
\hat{\dot{\mathbf A}}_{T,\vec k}(t) \cdot \hat{\dot{\mathbf
A}}_{T,-\vec k}(t)+ k^2~\hat{{\mathbf A}}_{T,\vec k}(t)\cdot
\hat{{\mathbf A}}_{T,-\vec k}(t) \right\} -\mathcal{C}_k
\label{numberop}
\end{equation}
\noindent where the index $\lambda$ labels the two independent
transverse polarization states and according to asymptotic theory of
interacting fields, $Z$ is identified with the wave function
renormalization for \emph{asymptotic states}, namely $Z$ is the
wave-function renormalization constant in the \emph{vacuum}. The
constant $\mathcal{C}_k$ will be adjusted so as to subtract the
photon number in the vacuum state. In free field theory
$Z=1\,,\,\mathcal{C}_k=1$. In asymptotic theory the field
$\hat{{\mathbf A}}_{T,\vec k}$ creates a single  particle state of
momentum $k$ with amplitude $\sqrt{Z}$ out of the vacuum state. The
quantities $Z\,, \mathcal{C}_k$ account for renormalization aspects
in the definition of the particle number in an interacting field
theory. A detailed study in  ref. \cite{nuestro} shows that once $Z$
and $\mathcal{C}_k$ are fixed by the wave function renormalization
constant and the normal ordering subtraction in the vacuum, these
terms cancel the zero temperature (vacuum) contributions. Therefore
this definition of the number operator does indeed describe the
number of photons \emph{in the medium} at time $t$.

Introducing  the real-time  correlation functions
\begin{eqnarray}
&& \langle A^+_{T,j}(\vec k;t) A^+_{T,l}(-\vec k;t') \rangle
=\mathcal{P}^{j,l}(\hat{\mathbf{k}})\left[
g^>_k(t,t') \Theta(t-t')+g^<_k(t,t')\Theta(t'-t)\right] \label{timeord}\\
&& \langle A^-_{T,j}(\vec k;t) A^-_{T,l}(-\vec k;t') \rangle
=\mathcal{P}^{j,l}(\hat{\mathbf{k}})\left[
g^>_k(t,t') \Theta(t'-t)+g^<_k(t,t')\Theta(t-t') \right]\label{antitimeord}\\
 &&\langle A^-_{T,j}(\vec k;t) A^+_{T,l}(-\vec k;t') \rangle  =  \mathcal{P}^{j,l}(\hat{\mathbf{k}})~g^>_k(t,t')
\label{phigreat} \\
 &&\langle A^-_{T,l}(-\vec k;t') A^+_{T,j}(\vec k;t)
\rangle  = \mathcal{P}^{j,l}(\hat{\mathbf{k}})~ g^<_k(t,t')
\label{philess}
\end{eqnarray}

\noindent where the superscripts $\pm$ refer to the forward $(+)$
and backwards $(-)$ time branches. The expectation value of this
number operator is defined  in eq. (\ref{totave}), and assuming
$\overline{{\mathcal{A}}^a_{i,\vec k}}=0;
\overline{{\mathcal{E}}^b_{i,-\vec k}}=0$ its expectation value in
the time evolved density matrix is found to be given by
\begin{equation} (2\pi)^3 \frac{dN}{d^3kd^3x} \equiv
\sum_{\lambda}\langle \hat{N}_{k,\lambda}(t) \rangle =
\frac{1}{2k\,Z} \left(\frac{\partial}{\partial t}
\frac{\partial}{\partial t'}+ k^2\right)
\left[g^>_k(t,t')+g^<_k(t,t')\right]|_{t=t'} -\mathcal{C}_k
\label{expnumb}
\end{equation}

In terms of the center of mass field $\vec{\mathcal{A}} $
introduced in eq. (\ref{wigvars}) it is straightforward to  find
that the correlation function in the bracket in eq.
(\ref{numberop}) is given by
\begin{equation}
\langle \mathcal{A}^a_{\vec k}(t) \mathcal{A}^b_{-\vec k}(t')
\rangle = \frac{\mathcal{P}^{ab}(\hat{\mathbf{k}})}{2}
\left[g^>_k(t,t')+g^<_k(t,t') \right] \label{CMcorre}
\end{equation}
and the occupation number can be written in terms of the center of
mass Wigner variable as follows
\begin{equation}
\sum_{\lambda}\langle \hat{N}_{k,\lambda}(t) \rangle = \frac{1}{2k
Z}\left[ \langle \dot{\vec{\mathcal{A}}}_{\vec k}(t) \cdot
\dot{\vec{\mathcal{A}}}_{-\vec k}(t)\rangle +k^2 \langle
\vec{\mathcal{A}}_{\vec k}(t) \cdot\vec{\mathcal{A}}_{-\vec
k}(t)\rangle\right]-\mathcal{C}_k \label{CMnumber}
\end{equation}

\noindent where the expectation values are obtained as in eq.
(\ref{totave}) and $ \vec{\mathcal{A}}_{\vec k}(t)$ is the
solution of the Langevin equation given by eq.
(\ref{inhosolution}).

Assuming for simplicity that $\overline{{\mathcal{A}}^a_{i,\vec
k}}=0; \overline{{\mathcal{E}}^b_{i,-\vec k}}=0$, the expectation
value of the number operator eq. (\ref{numberop})  as defined by
eq. (\ref{totave}) is given by

\begin{eqnarray}
(2\pi)^3 \frac{dN}{d^3kd^3x} \equiv \sum_{\lambda}\langle
\hat{N}_{k,\lambda}(t) \rangle & = & \frac{1}{k~Z}\left\{
\frac{1}{2k}\left[1+2\mathcal{N}_k\right]\left[ \ddot{f}^2_k(t) + 2 k^2~ \dot{f}^2_k(t) + k^4 ~ f^2_k(t) \right] \right. \nonumber \\
&  & \left. + \int_{-\infty}^{\infty} \frac{d\omega}{2\pi}
\tilde{\mathcal K}_T(k,\omega) \left[ |\mathcal{F}_k(\omega,t)|^2
+ k^2 |\mathcal{H}_k(\omega,t)|^2 \right] \right\} -\mathcal{C}_k
\label{finumber}
\end{eqnarray}

\noindent where we have introduced the auxiliary functions

\begin{eqnarray}
&& \mathcal{H}_k(\omega,t)= \int_0^t d\tau {f}_k(\tau)e^{-i\omega
\tau}\label{funH}\\
 && \mathcal{F}_k(\omega,t)= \int_0^t d\tau
\dot{f}_k(\tau)e^{-i\omega \tau}\label{funF}
\end{eqnarray}
\noindent and $\tilde{\mathcal K}_T(k,\omega)$ is given by the
fluctuation-dissipation relation eq. (\ref{flucdiss}) and
$\mathcal{N}_k$ is the initial photon distribution function.

This is one of the main results of this study. The expression
(\ref{finumber}) is truly \emph{non-perturbative} since the
function $f_k(t)$ given by (\ref{foft}) describes the real time
evolution after a \emph{Dyson} resummation of the photon
propagator in terms of the geometric series with the polarization
$\Sigma$ to lowest order in $\alpha_{em}$ but in principle to all
orders in $\alpha_s$.

Thus it is clear that the  formulation in terms of the
non-equilibrium effective action for the physical transverse gauge
fields described above leads to a novel non-perturbative approach
to study photon production from a thermal source directly in real
time.

\section{Lowest order in perturbation theory:}\label{loword}

The results obtained above are general and as such may be
unfamiliar within the context of  photon production from a QGP in
LTE. In order to establish the relationship to the usual approach
we now obtain the photon yield in strict perturbation theory,
namely to lowest order in $\alpha_{em}$. The Laplace transform
(\ref{laplaf}) to lowest order in the perturbative expansion in
$\alpha_{em}$ is given by

\begin{equation}\label{Laplapert}
\tilde{f}_k(s) = \frac{1}{s^2+k^2} -
\frac{\tilde{\Sigma}_T(k;s)}{(s^2+k^2)^2}+\mathcal{O}(\alpha^2_{em})
\end{equation}

The fundamental solution can be readily obtained by inverting the
Laplace transform using eq. (\ref{laplasig})  and is found to be

\begin{equation}
f_k(t) = \frac{\sin[kt]}{k}+\delta
f_k(t)+\mathcal{O}(\alpha^2_{em})
\end{equation}
\noindent with

\begin{eqnarray}\label{deltaf}
\delta f_k(t) = &&\frac{\sin[kt]}{k}~\int^{\infty}_{-\infty}
\frac{d\omega}{\pi}\mathrm{Im}\tilde{\Sigma}^R_T(k;\omega)
\Bigg\{\frac{1}{2k^2(\omega-k)}-\frac{1}{2k(\omega-k)^2}\Bigg\}-\frac{t\cos[kt]}{k}~
\int^{\infty}_{-\infty}
\frac{d\omega}{\pi}\frac{\mathrm{Im}\tilde{\Sigma}^R_T(k;\omega)}{2k(\omega-k)}\nonumber
\\ && + \int^{\infty}_{-\infty}
\frac{d\omega}{\pi}{\mathrm{Im}\tilde{\Sigma}^R_T(k;\omega)}\frac{\sin[\omega
t]}{(\omega^2-k^2)^2}
\end{eqnarray}

Inserting this perturbative solution in eq. (\ref{finumber}) and
keeping terms consistently up to $\mathcal{O}(\Sigma_T)\sim
\mathcal{O}(\alpha_{em})$ and setting $\mathcal{N}_k=0$ we find to
lowest order in $\alpha_{em}$

\begin{equation}\label{OralfaN}
(2\pi)^3 \frac{dN(t)}{d^3kd^3x} =
\left[\frac{1}{Z}-\mathcal{C}_k\right]+\frac{1}{k}\int^{\infty}_{-\infty}
\frac{d\omega}{\pi}{2\,\mathrm{Im}\tilde{\Sigma}^R_T(k;\omega)}\,n(\omega)\,
\frac{1-\cos[(\omega-k)t]}{(\omega-k)^2}
\end{equation}

The $1/Z$ in the first term is a consequence of introducing the
wave function renormalization  in the definition of the photon
number in equation (\ref{numberop}), setting $Z=1;\mathcal{C}_k=1$
corresponding  to the free field expression, the first term
vanishes. The second term is \emph{exactly} the one obtained in a
perturbative expansion in \emph{real time} in reference
\cite{nuestro}. This equivalence can be inferred by noticing that
the transverse part of the polarization $\Pi_T(k,\omega)$ as
defined in eq. (IV.9) in reference \cite{nuestro} and
$\tilde{\Sigma}^R_T(k;\omega)$ as defined by eq. (\ref{polasplit})
above imply that $\Pi_T(k,\omega)\equiv
2\tilde{\Sigma}^R_T(k;\omega)$. Of course the perturbative study
of reference \cite{nuestro} defined the number of photons as in
free field theory, corresponding to setting $Z=1;\mathcal{C}_k=1$.
Thus we see that the perturbative evaluation of the expression
(\ref{CMnumber}) to lowest order in $\alpha_{em}$ is exactly the
same as obtained in the real time study in \cite{nuestro}.

Furthermore, as emphasized in ref.\cite{nuestro} the asymptotic
long time limit takes the form
\begin{equation}
\frac{d N(t)}{d^3x d^3k} = \frac{1}{(2\pi)^3\,k}
\left[\frac{2\,\mathrm{Im}\tilde{\Sigma}^R_T(k;\omega=k)}{e^{\frac{k}{T}}-1}
\; t+\int^{+\infty}_{-\infty}\frac{d\omega}{\pi}
\frac{2\,\mathrm{Im}\tilde{\Sigma}^R_T(k;\omega)}{e^{\frac{\omega}{T}}-1}
\; \mathcal{P}\frac{1}{(\omega-k)^2}+\mathcal{O}\left(\frac{1}{t}
\right) \right]\label{longtimeyield} \; .
\end{equation}

Therefore in the asymptotic long time limit and to lowest order in
$\alpha_{em}$ the \emph{photon production rate}

\begin{equation}
\frac{d N(t)}{d^4x d^3k} = \frac{1}{(2\pi)^3}
\frac{2\,\mathrm{Im}\tilde{\Sigma}^R_T(k;\omega=k)}{k(e^{\frac{k}{T}}-1)}.\label{Srate}
\end{equation}

\noindent is \emph{exactly} the S-matrix result\cite{nuestro}. In
the case that $\mathrm{Im}\tilde{\Sigma}^R_T(k;\omega=k)=0 $ as
for example in the hard thermal loop
approximation\cite{brapis,kapustabook,lebellac}  a logarithmic
dependence replaces the linear time growth in eq.
(\ref{longtimeyield})\cite{wangphoton,nuestro}.

\subsection{Photon production from
kinetics:}\label{subsec:kinetics}
Kinetic theory provides an alternative approach to photon
production. Within a simple kinetic description the  time
evolution of the number of photons in a phase space cell is given
by a gain minus loss (master) type kinetic equation, which  for
the distribution function of each polarization is given by

\begin{equation}
\frac{d~n_{k,\lambda}(t)}{dt} = [1+n_{k,\lambda}(t)]\Gamma^>_k -
n_{k,\lambda}(t) \Gamma^<_k \label{kinetic}
\end{equation}
\noindent with the photon number (assuming an isotropic
distribution) per polarization

\begin{equation}
n_{k,\lambda}(t) = (2\pi)^3 \frac{d^3N_{\lambda}}{d^3xd^3k}
\end{equation}

\noindent and $\Gamma^>_k~;~\Gamma^<_k$ are the forward and
backward rates which are computed using Fermi's Golden rule which
is equivalent to the S-matrix calculation  of transition rates.
Detailed balance entails

\begin{equation}\label{detbal}
\Gamma^>_k = e^{-\beta k}\, \Gamma^<_k
\end{equation}

The solution of the kinetic equation (\ref{kinetic}) with the
initial condition $n_{k,\lambda}(0)=0$ gives the following time
evolution for the sum over the polarization

\begin{equation}
n_{k}(t) = 2n_{eq}(k)(1-e^{-\gamma_k t})~~;~~ \gamma_k =
\Gamma^<_k-\Gamma^>_k \label{solkin}
\end{equation}
\noindent with $n_{eq}(k)$ the equilibrium Bose-Einstein
distribution function for photons.

This solution illuminates two important aspects: i) the
thermalization time scale is $\tau_k = 1/\gamma_k$ and ii) for
$t<< \tau_k $

\begin{equation}
n_{k}(t) = 2 n_{eq}(k)\, \gamma_k \, t +
\mathcal{O}(\gamma^2_k~t^2)
\end{equation}

\noindent the photon production rate is precisely determined by
the expression for the photon number during time scales much
shorter than that for thermalization, namely

\begin{equation}\label{phoprodkin}
\frac{d~n_{k}(t)}{dt} = 2 n_{eq}(k) \gamma_k
\end{equation}

The relaxation rate of the distribution function  $\gamma_k=
2\Gamma_k$ with $\Gamma_k$ relaxation (damping) rate of the
single-particle excitation, in this case the photon damping rate,
which  is given by

\begin{equation}\label{gammasig}
\Gamma_k = \frac{\mbox{Im}\tilde{\Sigma}_T(k,\omega=k)}{2 k}
\end{equation}
\noindent where the factor $k$ in the denominator refers to the
free photon mass shell. Accounting for the two transverse
polarization state, one obtains the rate of photon production

\begin{equation}\label{rateSm}
\frac{d~n_k(t)}{dt} =
\frac{2\,\mbox{Im}\tilde{\Sigma}_T(k,\omega=k)}{k(e^{\frac{k}{T}}-1)}
\end{equation}

\noindent which is the result obtained above in eq. (\ref{Srate}).
This simple kinetic description clearly shows the main physical
assumptions and ingredients that enter in the computation of the
photon production rate.

The  kinetic equation (\ref{kinetic}) is obtained from a full
quantum field theory description by defining the photon number
operator as in the free field quantum theory of photons, thus the
time evolution of this operator is solely determined by the
interaction\cite{boyanphoton}. The expectation value of the number
operator is obtained in a perturbative expansion using the real
time Feynman rules with free field photon propagators that include
the distribution function just as in the equilibrium
case\cite{boyanphoton}. The main point of this discussion is that
in the quantum field theory formulation leading to the simple
kinetic equation (\ref{kinetic}) the photons propagate as
\emph{free particles}, namely with the free photon dispersion
relation. This ignores the fact that in the medium photons
propagate as collective modes, \emph{not} as single particle
excitations. In order to include the collective effects in the
medium, which are more relevant for photons with momenta smaller
than or of the order of the
temperature\cite{brapis,kapustabook,lebellac}, a non-perturbative
description is required.

\subsection{Kinetic interpretation of the S-matrix
result:}\label{subsec:kinSmtx}

The analysis of the previous section leads to a kinetic
interpretation of the S-matrix result. The S-matrix approach
extracts the forward rate $\Gamma^>_k$ by computing the transition
probability in the infinite time limit, assuming $n(k)=0$ and
\emph{ignoring the build up of the photon population}, namely the
terms with $n_{k,\lambda}$ in the kinetic equation
(\ref{kinetic}), thus leading to

\begin{equation}\label{Smat}
\frac{dn_k}{dt} = 2 \Gamma^>_k
\end{equation}

Using the detailed balance relation eq. (\ref{detbal}) and the
relation $\gamma_k = 2\Gamma_k$ with $\Gamma_k$ given by eq.
(\ref{gammasig}) one finds that eq. (\ref{Smat}) is equivalent to
the result (\ref{rateSm}).

Obviously the buildup of the population can only be neglected
during a time scale $t \ll 1/\Gamma_k$, however the S-matrix
approach manifestly takes the time to infinity extracting only the
terms that grow linearly in time in this limit, and  ignoring
terms that grow slower and that can make important contributions
during a finite time scale\cite{wangphoton,nuestro}. In
\emph{assuming} that the S-matrix approach is therefore valid only
during a finite interval of time before the photon population
builds up highlights the inconsistency of neglecting time
dependent contributions associated with the finite lifetime. While
this point has been emphasized in refs.\cite{wangphoton,nuestro}
the analysis based on the kinetic equation makes it  explicit.

A kinetic interpretation of the S-matrix result as gleaned from
the full non-perturbative solution of the kinetic equation
(\ref{solkin}) first assumes no initial population, that the rate
is constant in time and that the population does not build up,
leading to

\begin{equation}\label{kinS}
\frac{dN_{SM}}{d^4x d^3k}=\left.\frac{dN_{kin}}{d^4x
d^3k}\right|_{t=0}= 2 \Gamma^>_k
\end{equation}

\noindent with the factor $2$ accounting for the two polarization
states.

\section{Non-perturbative aspects I: Breit-Wigner (narrow width) approximation:}

Having confirmed that the lowest order in the strict perturbative
expansion in $\alpha_{em}$ of the full expression (\ref{finumber})
coincides with the results previously obtained in the literature,
we now proceed to study non-perturbative aspects. The first stage
of our study is to make contact with the kinetic approach to
photon production studied in section \ref{subsec:kinetics} above.

For weak couplings ($\alpha_{em}$ \emph{and} $\alpha_s$) when the
width of the quasiparticle is much smaller than its energy, the
photon spectral density

\begin{equation}
\rho_{\gamma}(k;\omega) = \frac{1}{\pi}
\frac{\left[\mathrm{Im}\tilde{\Sigma}^R_T(k;\omega)\right]}{\left[\omega^2-k^2-\mathrm{Re}\tilde{\Sigma}^R_T(k;\omega)
\right]^2+\left[\mathrm{Im}\tilde{\Sigma}^R_T(k;\omega)\right]^2}\label{specphot}
\end{equation}

\noindent features a pole in the second (unphysical) Riemann sheet
but near the real axis at the position of the ``quasiparticle''
pole, which is a solution of the equation

\begin{equation}
\omega^2_p(k)-k^2-\mathrm{Re}\tilde{\Sigma}^R_T(k;\omega_p(k))=0
\label{quasipole}
\end{equation}

The imaginary part of the transverse photon self-energy evaluated
at $\omega=\omega_p(k)$ determines the width or damping rate of
the single quasiparticle excitation. If
$\mathrm{Im}\tilde{\Sigma}^R_T(k;\omega_p(k))=0$ then the
quasiparticle is stable corresponding to a true pole in the
physical sheet. If
$\mathrm{Im}\tilde{\Sigma}^R_T(k;\omega_p(k))\neq 0$ the
quasiparticle pole moves off the physical sheet becoming a
resonance. In weak coupling the spectral density can be
approximated by a Breit-Wigner form near the quasiparticle pole

\begin{equation}\label{BW}
\rho_{\gamma}(k;\omega\thickapprox\omega_p(k)) =
\frac{\mathcal{Z}_k}{2\pi\omega_p(k)}~ \frac{\Gamma_k}{(\omega
-\omega_p(k))^2+\Gamma^2_k}
\end{equation}

\noindent with the residue at the quasiparticle pole and the width
given by

\begin{equation}\label{reswidth}
\mathcal{Z}^{-1}_k = 1- \frac{\partial\,
\mathrm{Re}\tilde{\Sigma}^R_T(k;\omega)}{\partial
\omega^2}\Bigg|_{\omega=\omega_p(k)}~~;~~
\Gamma_k=\mathcal{Z}_k\frac{\mathrm{Im}\tilde{\Sigma}^R_T(k;\omega_p(k))}{2\omega_p(k)}
\end{equation}

The residue at the quasiparticle pole $\mathcal{Z}_k$ is in
principle different from $Z$ the wave function renormalization
that enters in the asymptotic theory definition of the photon
number (\ref{numberop}). The reason for the difference is that the
photons that are measured in the detector are asymptotic states,
hence $Z$ in this definition should refer to the \emph{vacuum}
value and not the in-medium residue at the quasiparticle pole.

In strict perturbation theory the connection with the kinetic
approach which applies to the distribution function of photons of
dispersion relation $\omega_p(k)=k$ is established by taking $Z=1;
\mathcal{Z}_k=1\,;\omega_p(k)=k\;,\mathcal{C}_k=1$, assuming that
there are no initial photons present, namely $\mathcal{N}_k=0$,
and assuming the Breit-Wigner form of the spectral density
(\ref{BW}) in the full range of $\omega$, not just near the
quasiparticle pole. Under these assumptions

\begin{equation}\label{BWpert}
\rho_{\gamma}(k;\omega) = \frac{1}{2\pi k}~
\frac{\Gamma_k}{(\omega
-k)^2+\Gamma^2_k}~~;~~\Gamma_k=\frac{\mathrm{Im}\tilde{\Sigma}^R_T(k;\omega=k)}{2k}
\end{equation}

In the narrow width approximation $\Gamma_k \ll k$ the fundamental
solution is given by

\begin{equation} \label{longti}
f_k(t) = \frac{\sin[kt]}{k}e^{-\Gamma_k t}
\end{equation}

Inserting this solution and neglecting terms of
$\mathcal{O}(\Gamma^2_k)$ in the numerator in eq.
(\ref{finumber}), and using the fluctuation-dissipation relation
(\ref{flucdiss}) we find

\begin{equation}\label{BWnum}
(2\pi)^3 \frac{dN(t)}{d^3kd^3x}   = e^{-2\Gamma_k t}-1+
\frac{1}{2k}
 \int_{-\infty}^{\infty} \frac{d\omega}{\pi}
 \frac{[1+2n(\omega)]~\mathrm{Im}\tilde{\Sigma}^R_T(k;\omega)}{(\omega-k)^2+\Gamma^2_k}\left[
 1+e^{-2\Gamma_k t}-2 e^{-\Gamma_k t}\cos[(\omega-k)t]  \right]
\end{equation}

In the weak coupling limit, assuming a narrow Breit-Wigner
spectral density, namely taking
$\mathrm{Im}\tilde{\Sigma}^R_T(k;\omega)\sim 2k \, \Gamma_k$ and
using that the spectral density is sharply peaked at $\omega \sim
k$ we can replace $n(\omega) \equiv n_{eq}(k)$ with $n_{eq}(k)$
the Bose-Einstein distribution function for photons. The integrals
can be done straightforwardly and we find

\begin{equation}\label{BWnumfin}
(2\pi)^3 \frac{dN(t)}{d^3kd^3x}   = 2 n_{eq}(k)(1-e^{-2\Gamma_k
t})
\end{equation}

This result is the same as the solution of the kinetic equation
(\ref{solkin}) since $\gamma_k = 2\Gamma_k$. The assumptions
leading to eq. (\ref{BWnumfin}), namely weak coupling, neglecting
self-energy corrections to the dispersion relation as well as
renormalization effects (wave-function renormalization) and the
narrow-width-Breit-Wigner approximation for the spectral density
are all approximations invoked in the Fermi's golden rule approach
to the kinetic description.

We note that even under all these assumptions, the result eq.
(\ref{BWnumfin})  is truly \emph{non-perturbative} and a result of
the non-perturbative Dyson-like resummation of the photon
propagator manifest in the function $f_k(t)$. The strict
perturbative evaluation as presented in the previous section
reveals that when $\mathrm{Im}\tilde{\Sigma}(k,\omega\sim k) \neq
0$, the photon yield grows linearly in time, a consequence of
Fermi's golden rule (or secular terms). Such strict perturbative
evaluation clearly is restricted to a time interval $t \ll
\Gamma^{-1}_k$ since for $t \gg \Gamma^{-1}_k$ the photon yield
attains the equilibrium form given by the Bose-Einstein
distribution function. The secular terms (terms that grow in time)
that appear in the strict perturbative expansion of eq.
(\ref{BWnumfin}) are typically manifest as pinch singularities in
finite temperature field theory\cite{pinch}. A dynamical
renormalization group program has been recently introduced that
provides a resummation of these secular terms and leads to their
exponentiation \cite{boyankinetic}. Thus, making contact with the
dynamical renormalization group resummation of terms that grow in
time, it is clear that the formulation presented above leads to a
resummation of the perturbative series. The description of photon
production given by our result eq. (\ref{finumber}) not only
coincides with the perturbative result in the strict perturbative
expansion, but provides a systematic Dyson-like resummation of the
perturbative series leading to a uniform expansion in the coupling
valid at all times. This resummation is similar to that implied by
the dynamical renormalization group approach of
ref.\cite{boyankinetic} and reveals the physics of the relaxation
of the population towards equilibrium as well as the corresponding
time scales.

\section{Non-perturbative aspects II: photons or
plasmons?}\label{sec:plasmon}

The discussion in the previous two sections above  confirmed that
the real time description of photon production describes both the
lowest order results known previously and is capable of describing
the thermalization of photons.

We now study non-perturbative aspects of the production and
\emph{propagation} of photons in a locally equilibrated QGP.

An important result from the hard thermal loop program in finite
temperature gauge field theories\cite{brapis,lebellac,kapustabook}
is that even for weak coupling the perturbative expansion must be
resummed for momenta $k \leq \omega_{pl}$ with $\omega_{pl}$ the
plasmon mass. To lowest order in the HTL program for $N_c=3$ and
two (massless, up and down) quarks, the plasmon mass is given by

\begin{equation}\label{plasmonmass}
\omega_{pl}= \sqrt{\frac{5}{27}}~ eT = 13.033
\left(\frac{T}{100~\mbox{Mev}}\right)~\mbox{Mev}
\end{equation}
\noindent which for example for the temperature expected at RHIC
$T \sim 300~\mathrm{Mev}$  $\omega_{pl} \sim 40~\mathrm{Mev}$.

The real and imaginary parts of the transverse photon polarization
in the HTL approximation are given by

\begin{eqnarray}
\mathrm{Re} \tilde{\Sigma}^R_T(k;\omega) & = & k^2 ( x^2-1 )-
\frac{3 \omega^2_{pl}}{2}\Bigg[x^2+
\frac{1}{2}x(1-x^2)\ln\left|\frac{x+1}{x-1}\right| \Bigg]~~;~~ x
\equiv \frac{\omega}{k}
\label{realHTL}\\
\mathrm{Im}\tilde{\Sigma}^R_T(k;\omega) & = &
\frac{3\pi}{4}\omega^2_{pl}~\frac{\omega}{k}\left(1-\frac{\omega^2}{k^2}
\right)\Theta(k^2-\omega^2) \label{imagHTL}
\end{eqnarray}

In the HTL approximation, the position of the plasmon pole as a
function of momentum $k$ is determined by the solution(s) of the
trascendental equation

\begin{equation}
x^2_p-1 - \frac{3 \omega^2_{pl}}{2k^2}\Bigg[x^2_p+
\frac{1}{2}x_p(1-x^2_p)\ln\Bigg[\frac{x_p+1}{x_p-1} \Bigg]
\Bigg]=0~~;~~ x_p= \frac{\omega_p(k)}{k}
\end{equation}

\begin{figure}[htbp]
\begin{center}
\epsfig{file=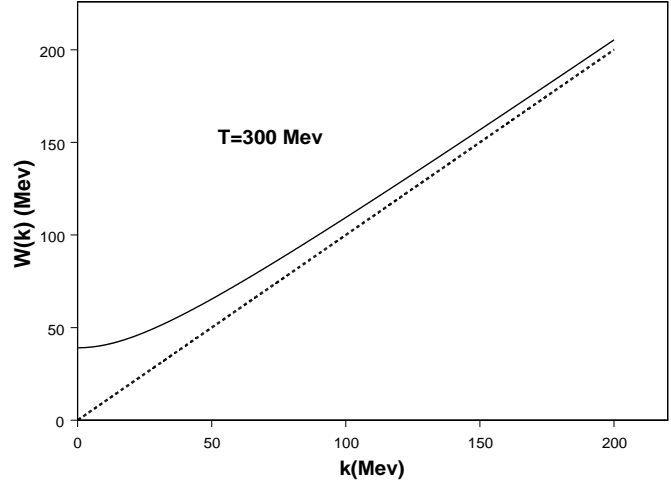,width=4in,height=3in}
 \caption{The  dispersion relation $\omega_p(k)$ (in Mev) vs. $k$ (in Mev/c) in the HTL approximation for
 $ \omega_{pl}= 39.099~\mathrm{Mev}$, corresponding to a
 temperature $T= 300~\mathrm{Mev}$ . The solid line is the plasmon and the dashed line the free photon dispersion
 relations respectively. } \label{fig:disprel}
 \end{center}
\end{figure}

The numerical solution for the dispersion relation is shown in
fig. (\ref{fig:disprel}) for $T\sim 300~ \mathrm{Mev}$
corresponding to a (transverse) plasma frequency
$\omega_{pl}=39.099 ~\mathrm{Mev}$.

Even for momenta $k\sim 200~\mathrm{Mev} \gg \omega_{pl}$ the
difference between the plasmon and the free particle dispersion
relation is about $15\% $. The real time solution $f_k(t)$ in the
HTL approximation is given by

\begin{eqnarray}\label{foftHTL}
f_k(t)& = & \mathcal{Z}_k
\frac{\sin[\omega_p(k)\,t]}{\omega_p(k)}+ f_c(k,t)\nonumber \\
f_c(k,t) & = & \int_{-k}^{k} \frac{d\omega}{\pi} \frac{\sin(\omega
t)~\mathrm{Im}\tilde{\Sigma}^R_T(k;\omega)}{\left[\omega^2-k^2-\mathrm{Re}\tilde{\Sigma}^R_T(k;\omega)
\right]^2+\left[\mathrm{Im}\tilde{\Sigma}^R_T(k;\omega)\right]^2}
\end{eqnarray}

\noindent where $\mathcal{Z}_k$ is the residue at the plasmon pole
(wavefunction renormalization) given by

\begin{equation}\label{residue}
\mathcal{Z}_k = \left[1- \frac{1}{2\omega_p(k)} \frac{\partial
\mathrm{Re}\tilde{\Sigma}_T(k;\omega)}{\partial
\omega}\Bigg|_{\omega=\omega_p(k)} \right]^{-1}
\end{equation}

The sum rule (\ref{sumrule}) entails the following

\begin{equation}\label{ressumrule}
\mathcal{Z}_k +\int_{-k}^{k} \frac{d\omega}{\pi}
\frac{\omega~\mathrm{Im}\tilde{\Sigma}^R_T(k;\omega)}{\left[\omega^2-k^2-\mathrm{Re}\tilde{\Sigma}^R_T(k;\omega)
\right]^2+\left[\mathrm{Im}\tilde{\Sigma}^R_T(k;\omega)\right]^2}
= 1
\end{equation}

The residue at the plasmon pole $\mathcal{Z}_k$ is shown as a
function of $k$ for $T\sim 300 ~\mathrm{Mev}$ in figure
(\ref{fig:Zofk}) below.

\begin{figure}[htbp]
\begin{center}
\epsfig{file=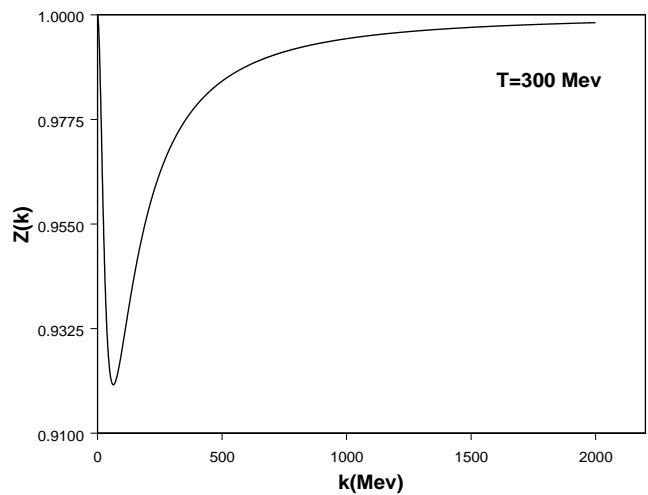,width=4in,height=3in}
 \caption{The residue at the plasmon (quasiparticle) pole  for $\omega_{pl}=39.099~\mathrm{Mev}$, corresponding to a
 temperature $T= 300~\mathrm{Mev}$ .} \label{fig:Zofk}
 \end{center}
\end{figure}

 Figure (\ref{fig:fcont})
below displays the continuum contribution $f_c(k,t)$ given by the
integral expression in eq. (\ref{foftHTL}) for the values
$k=50;100;150~\mathrm{Mev}/c$ respectively for $T\sim
300~\mathrm{Mev}$. It is clear from this figure that the continuum
contribution is perturbatively small, of
$\mathcal{O}(\alpha_{em})$ and damps out on time scales of $\sim
10 ~\mbox{fm}/c$ or longer even for  large momenta. This time
scale is of the order of the lifetime of the QGP expected at RHIC
or LHC, hence $f_c(k,t)$ will contribute to
$\mathcal{O}(\alpha_{em})$ during the lifetime of the QGP.

\begin{figure}[htbp]
\begin{center}
\epsfig{file=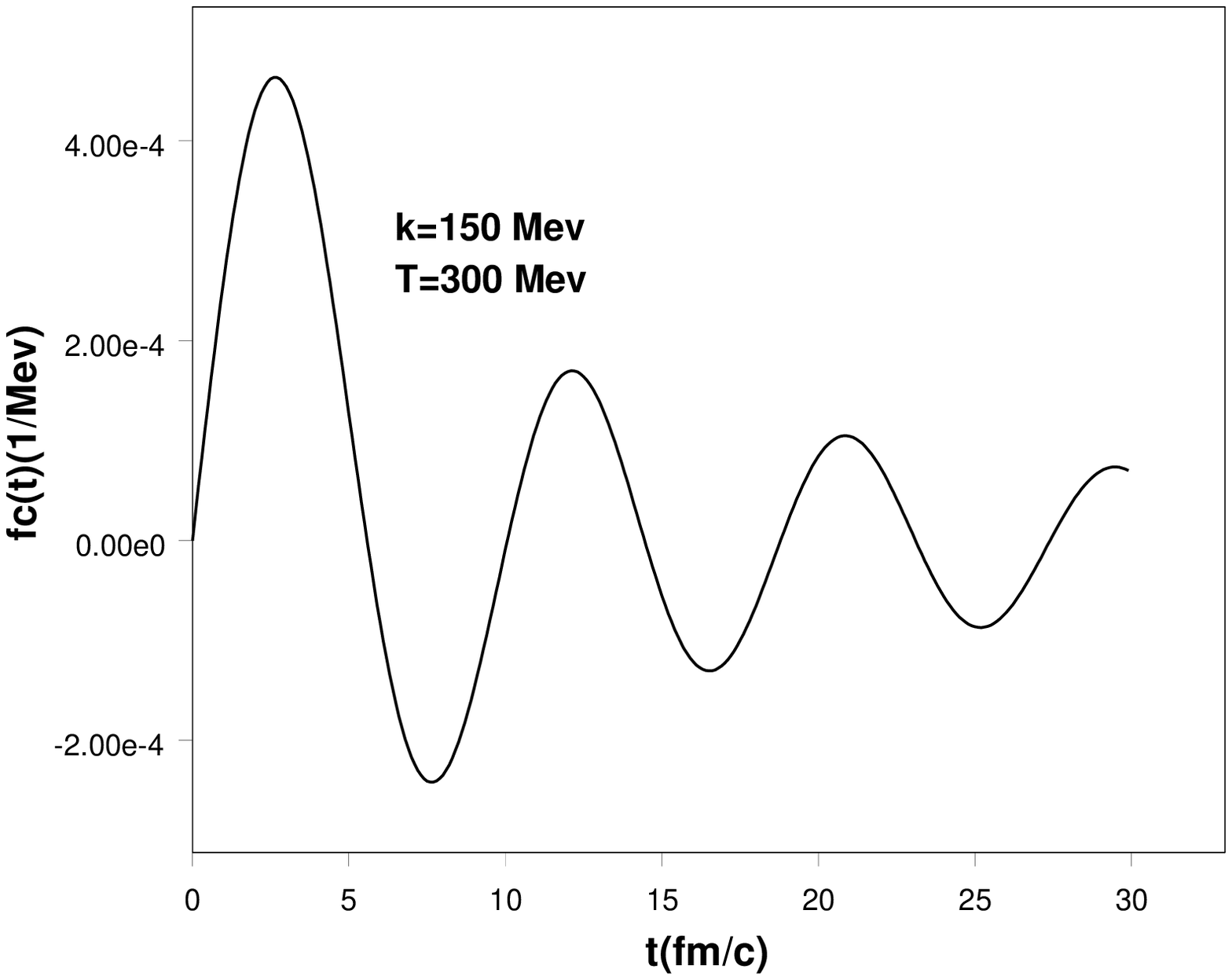,width=3in,height=3in}
\epsfig{file=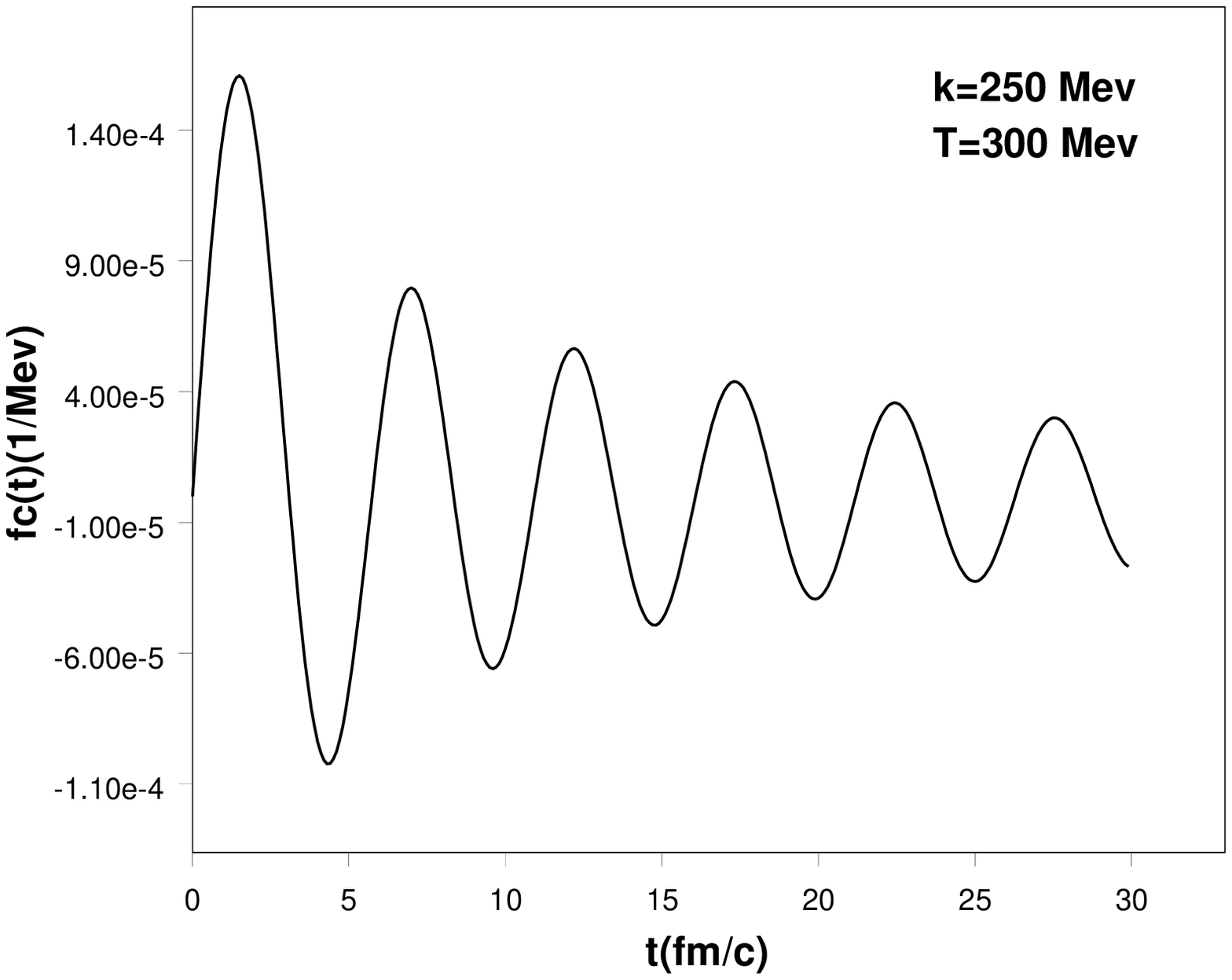,width=3in,height=3in}
\epsfig{file=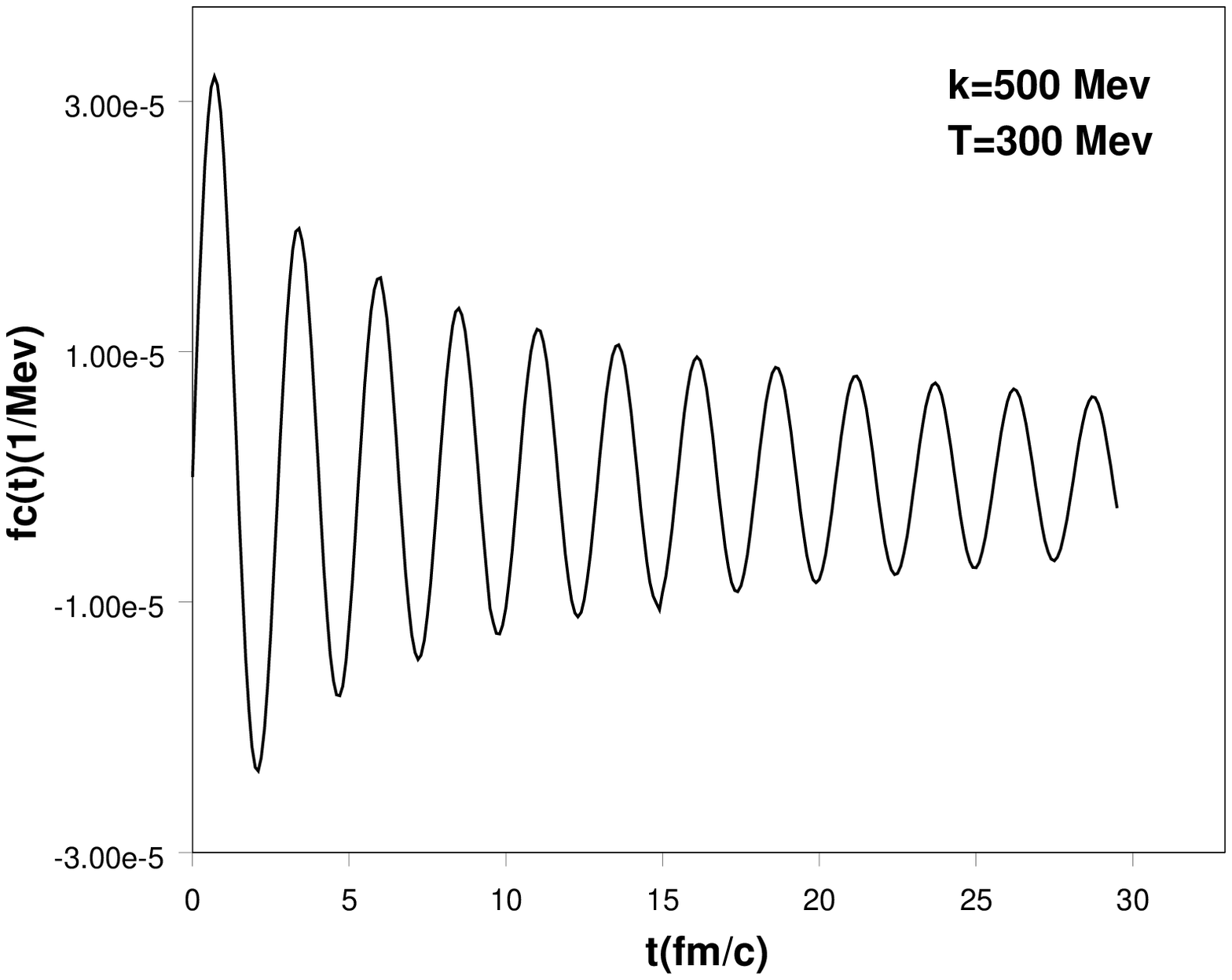,width=3in,height=3in}
 \caption{The continuum contribution $f_c(k,t)$ given by the integral in (\ref{foftHTL})
 for $k=150;250;500~\mathrm{Mev}$ respectively as a function of
 $t$, for  $T\sim 300~\mathrm{Mev}$ .} \label{fig:fcont}
 \end{center}
\end{figure}

The main point of this discussion is to highlight that once the
photon is produced at a space-time vertex, it propagates in the
medium as a \emph{plasmon} collective mode, not as a free
electromagnetic plane wave. The propagation of the photon in the
medium as a collective mode has obviously nothing to do with the
thermalization of the produced photon and only depends on the
wavelength of the photon being smaller than the spatial size of
the system. The situation is similar to an electromagnetic wave
propagating in a dispersive medium with an index of refraction,
the mean free path of the photon in the medium can be larger than
the spatial extension of the medium itself, but the
electromagnetic wave propagates (almost undamped) as a combination
of normal modes in the medium. This point has been advanced by
Weldon\cite{weldon} within the context of dilepton production in a
thermalized QGP\footnote{D.B. thanks Abhijit Majumder for
discussions on this point}.

This point can be simply illustrated by considering the average
over the noise of the solution (\ref{inhosolution}) with an
initial value of the expectation value of the transverse gauge
field and its time derivative corresponding to the case of a
positive energy plane wave, namely

\begin{equation}
\mathcal{A}^i_{\vec k,l}= \frac{\vec{\epsilon}_{\vec
{k},l}(\lambda)}{\sqrt{2k}}~~;~~ \mathcal{E}^i_{\vec k,l}= -ik
\,\mathcal{A}^i_{\vec k,l}\label{inicondplanewave}
\end{equation}
\noindent with $\vec{\epsilon}_{\vec {k},l}(\lambda)$ being a
transverse polarization vector.

The propagation of this initial state is given by eq.
(\ref{inhosolution}). The noise average eq. (\ref{noisecorrel})
implies

\begin{equation}
\langle \mathcal{A}_{\vec k,l}(t) \rangle = \vec{\epsilon}_{\vec
{k},l}(\lambda)\left[
e^{i\omega_p(k)t}~\frac{\mathcal{Z}_k}{2\sqrt{2k}}\left(1-\frac{k}{\omega_p(k)}\right)+
e^{-i\omega_p(k)t}~\frac{\mathcal{Z}_k}{2\sqrt{2k}}\left(1+\frac{k}{\omega_p(k)}\right)+\frac{1}{\sqrt{2k}}\dot{f}_c(k,t)-i\sqrt{\frac{k}{2}}
f_c(k,t) \right]\label{propa}
\end{equation}

Clearly  the propagation of the transverse electromagnetic wave is
not free. While the continuum contribution $f_c(k,t)$ vanishes at
long time, its presence guarantees that the initial conditions
(\ref{inicondplanewave}) are fulfilled via the sum rule
(\ref{ressumrule}).

\subsection{Beyond HTL}\label{beyo}

The real and imaginary parts of the (transverse) photon self
energy given by eqns. (\ref{realHTL}), (\ref{imagHTL}) are only
valid in the hard thermal loop approximation $k,\omega \ll T$. The
full one loop  expression for the finite temperature contribution
to the imaginary  part of the transverse photon self energy is
given by

\begin{equation}
\mathrm{Im}\tilde{\Sigma}^{(1)}_T(k;\omega)  =
\mathrm{Im}\tilde{\Sigma}^{(1)}_{LD}(k;\omega)
+\mathrm{Im}\tilde{\Sigma}^{(1)}_{2P}(k;\omega) \label{totimsig}\\
\end{equation}

\noindent where $\mathrm{Im}\tilde{\Sigma}^{(1)}_{LD}(k;\omega)
\,;\,\mathrm{Im}\tilde{\Sigma}^{(1)}_{2P}(k;\omega)$ are the
Landau damping and two particle cut respectively, given
by\cite{nuestro}

\begin{eqnarray}
 \mathrm{Im}\tilde{\Sigma}^{(1)}_{LD}(k;\omega)&  = & \frac{5}{3} \;
\alpha_{em}\,T^2 \left(1-\frac{\omega^2}{k^2}\right)\Bigg[
\frac{k}{T} \ln \left(\frac{1+e^{-W_-}}{1+e^{-W_+}}
\right)+\frac{2T}{k} \sum_{m=1}^{\infty}(-1)^{m+1}
\left(\frac{2}{m^3}+\frac{k}{T\,m^2} \right)\times\nonumber \\
&&  \left(e^{-m\,W_-}-e^{-m\,W_+} \right)
\Bigg]\Theta(k^2-\omega^2)\,\mathrm{sign}(\omega)  \label{imsigLD}\\
\mathrm{Im}\tilde{\Sigma}^{(1)}_{2P}(k;\omega)& = & \frac{5}{3} \;
\alpha_{em}\,T^2 \left(\frac{\omega^2}{k^2}-1\right)\Bigg[
\frac{k}{T} \ln \left(\frac{1+e^{-W_+}}{1+e^{-W_-}}
\right)-\frac{2T}{k} \sum_{m=1}^{\infty}(-1)^{m+1}
\Big[\frac{2}{m^3}\left(e^{-m\,W_-}-e^{-m\,W_+} \right)\nonumber\\
&& -\frac{k}{T\,m^2}\left(e^{-m\,W_-}+e^{-m\,W_+} \right)
\Bigg]\Theta(\omega^2-k^2)\,\mathrm{sign}(\omega) \; ,  \label{pi2P} \\
 W_{\pm}  & = &  \Big | \frac{|\omega|\pm k}{2T}\Big | \nonumber
\end{eqnarray}

\noindent and the superscript $(1)$ refers to one-loop.

The real part of the self energy is found by the dispersion
relation eq. (\ref{specsigret}). The two particle cut contribution
above the light cone would lead to an imaginary part for the
plasmon since the plasmon dispersion relation lies above the light
cone. This imaginary part is obviously related to the decay of the
plasmon into  \emph{bare massless} fermion-antifermion pairs.
Clearly this is unphysical, finite temperature self-energy
corrections to the fermion lines lead to a thermal mass for the
intermediate fermions which in the HTL approximation is at least
$22\%$ larger than the plasmon mass\footnote{Even much larger than
this estimate when the value for $\alpha_s$ given by eq.
(\ref{alfastrongofT}) is included through gluon contributions to
the fermion self energy.}, thus preventing plasmon decay into
fermion quasiparticles. The correct imaginary part of the
transverse photon self energy above the light cone requires a
non-perturbative resummation that involves not only fermion
self-energy corrections but also vertex corrections to satisfy the
Ward identities. While the damping rate for a non-abelian plasmon
at rest has been computed in ref.\cite{pisplasrate} and that for
the abelian counterpart was estimated for large $k$ in
(\cite{weldon}) the calculation of  the imaginary part of the
(transverse) photon self energy for all values of $\omega, k$ is
not yet available. What \emph{is} available in the literature is
the photon production rate calculated by the S-matrix approach to
lowest order in $\alpha_{em}$ and up to leading logarithmic order
in $\alpha_s$\cite{AMY}. From this result we can extract
${Im}\tilde{\Sigma}^R_T(\omega=k,k)$ by making use of the relation
eq. (\ref{Srate}).

The analysis presented above indicates that in order to obtain the
complete real time dependence of $f_k(t)$, hence the real time
photon yield, what is needed is the photon (transverse) self
energy for \emph{all values} of $\omega;k$. Below the light cone,
the Landau damping contribution (\ref{imsigLD}) is the leading
order result, of $\mathcal{O}(\alpha_{em} \alpha^0_s)$ whereas
above the light cone a resummation of the fermion lines and vertex
is required for a consistent analysis of the plasmon width. Since
the full imaginary part of the transverse photon self-energy above
the light cone is not available, only its value on the \emph{free
photon} mass shell, namely $\omega=k$, we \emph{model} the
imaginary part of the photon self energy in the full range of
frequency as

\begin{equation}\label{imamodel}
\mathrm{Im}\tilde{\Sigma}(k;\omega) \approx
\mathrm{Im}\tilde{\Sigma}^{(1)}_{LD}(k;\omega)+\mathrm{Im}\tilde{\Sigma}^{PP}(k;\omega=k)\,\Theta(\omega^2-k^2)\,\mathrm{sign}(\omega)
\end{equation}

\noindent where $\mathrm{Im}\tilde{\Sigma}^{PP}(k;\omega=k)$ is
extracted from the S-matrix photon production rate eq.
(\ref{Srate}) and the leading logarithmic result quoted in
ref.\cite{AMY}, resulting in the following expression for the
imaginary part on the \emph{free photon mass shell}

\begin{eqnarray}\label{imsigPP}
&&\mathrm{Im}\tilde{\Sigma}^{PP}(k;\omega=k)  = \frac{20 \pi
T^2}{9} \; \alpha_{em} \; \alpha_s(T) \;
\tanh\left[\frac{k}{2T}\right]\left[\ln\left(\frac{\sqrt{3}}{4\pi\alpha_s(T)}
\right)+C_{tot}\left(\frac{k}{T}\right)\right] \nonumber\\
&& C_{tot}(z)  =  \frac{1}{2}\ln(2z)+\frac{0.041}{z}-0.3615+1.01
\; e^{-1.35\,z}+\sqrt{\frac{4}{3}}
\left[\frac{0.548}{z^{\frac{3}{2}}} \, \ln\left(12.28+\frac{1}{z}
\right)+\frac{0.133 \; z}{\sqrt{1+\frac{z}{16.27}}} \right] \;
\end{eqnarray}

The strong coupling $\alpha_s$ is a function of temperature, we
will use the lattice parametrization \cite{karsch} for the
temperature dependence of the strong coupling $\alpha_s(T)$ given
by
\begin{equation}\label{alfastrongofT}
\alpha_s(T)= \frac{6\pi}{29 \ln\frac{8T}{T_c}}~~;~~T_c \sim
0.16~\mathrm{Gev} \; .
\end{equation}
Although this lattice fit is valid at high temperatures and
certainly not near the hadronization phase transition, we will
assume its validity in the temperature range relevant for RHIC in
order to obtain a numerical estimate of the photon yield. We note,
however, that at the temperature expected at RHIC or LHC $T \sim
300 -400 ~\mbox{Mev}$ $\alpha_s(0.3 ~\mathrm{Gev})\sim 0.24 $ and
the validity of the perturbative expansion in $\alpha_s$ (even
with leading logarithmic corrections) is at best questionable.

The assumption that the imaginary part is constant above the light
cone is consistent with the Breit-Wigner approximation. The
one-loop Landau damping contribution below the light cone leads to
the plasmon dispersion relation and is the leading order
($\mathcal{O}(\alpha_{em})$) contribution to the photon
self-energy. The damping of the plasmon excitation is \emph{not}
correctly described by this Breit-Wigner form, since the correct
plasmon damping rate is given by (see eq. (\ref{reswidth}) )

\begin{equation}\Gamma_k = \frac{\mathcal{Z}_k}{2\omega_p(k)} \;
\mathrm{Im}\tilde{\Sigma}_T[k,\omega=\omega_p(k)] \;
\end{equation}

However, since $\omega_p(k) $ differs from $k$ by less than $10\%$
for the relevant range $k\geq 200~\mbox{Mev}$, and \emph{assuming}
that $\mathrm{Im}\tilde{\Sigma}_T(k,\omega)$ is a smooth function
of $\omega$ near $\omega_p(k)$ it is reasonable to assume that
$\mathrm{Im}\tilde{\Sigma}_T(k,\omega=\omega_p(k))\sim
\mathrm{Im}\tilde{\Sigma}_T(k,\omega=k)$. A reliable  estimate of
the error incurred with this approximation requires  knowledge of
$\mathrm{Im}\tilde{\Sigma}_T(k,\omega)$ for all values of $\omega
\geq k$ near the plasmon pole which is not yet available and its
calculation is certainly beyond the scope of this article.

 Consistent with the above approximation for the imaginary part, the lowest order $\mathcal{O}(\alpha_{em})$
 contribution for the real part is obtained by the dispersion
 relation eq. (\ref{specsigret}). In order to guarantee the vanishing of the
 magnetic mass in the abelian theory\cite{lebellac}, we subtract the dispersion
 relation at zero frequency. The real part is therefore given by

 \begin{equation}\label{realsig1}
\mathrm{Re}\tilde{\Sigma}^{(1)}_T(k,k_0) = -\frac{2k^2_0}{\pi}
\int^k_0 d\omega~\mathcal{P}
\frac{\mathrm{Im}\tilde{\Sigma}^{(1)}_{LD}(k;\omega)}{\omega(\omega^2-k^2_0)}
 \end{equation}

\noindent where we have performed a subtraction at $k_0=0$ to
ensure the vanishing of the magnetic mass and used the property
(\ref{odd}), $\mathcal{P}$ stands for the principal part.

The spectral density eq. (\ref{specphot}) with the real and
imaginary parts of the self energy given by eq.(\ref{realsig1})
and (\ref{imamodel}), (\ref{imsigLD}),(\ref{imsigPP}) respectively
features a sharp peak above the lightcone at a value of the
frequency given by

\begin{equation}
\omega^2_p(k)-k^2-\mathrm{Re}\tilde{\Sigma}^{(1)}_T(k;k_0=\omega_p(k))=0
\end{equation}

We find numerically that the plasmon dispersion relation
$\omega_p(k)$ and residue at the plasmon pole $\mathcal{Z}_k$ are
remarkably similar to that obtained in the HTL approximation for
 momenta $k \geq 200 ~\mbox{Mev}$ for $300\leq T \leq 500 ~\mbox{Mev}$ displayed in figures
 (\ref{fig:disprel},\ref{fig:Zofk}) respectively.
This similarity was already pointed out in ref.\cite{thoma} and
our numerical study confirms the results obtained in that
reference.  This fact can be understood easily since for $k \ll T$
the full one loop imaginary part reduces to that in the HTL
approximation, and for $k \gg  T$ the corrections are truly
perturbative. Above the light cone the photon spectral density can
be very well approximated by the Breit-Wigner form

\begin{equation}\label{BWphot}
\rho_{\gamma}(k;\omega> k) =
\frac{\mathcal{Z}_k}{2\pi\omega_p(k)}~ \frac{\Gamma_k}{(\omega
-\omega_p(k))^2+\Gamma^2_k}
\end{equation}

\noindent with the residue at the plasmon pole and the width given
by

\begin{equation}\label{reswidth1}
\mathcal{Z}^{-1}_k = 1- \frac{\partial\,
\mathrm{Re}\tilde{\Sigma}^{(1)}_T(k;\omega)}{\partial
\omega^2}\Bigg|_{\omega=\omega_p(k)}~~;~~
\Gamma_k=\mathcal{Z}_k\frac{\mathrm{Im}\tilde{\Sigma}^{PP}(k;\omega=k)}{2\omega_p(k)}
\end{equation}

\begin{figure}[htbp]
\begin{center}
\epsfig{file=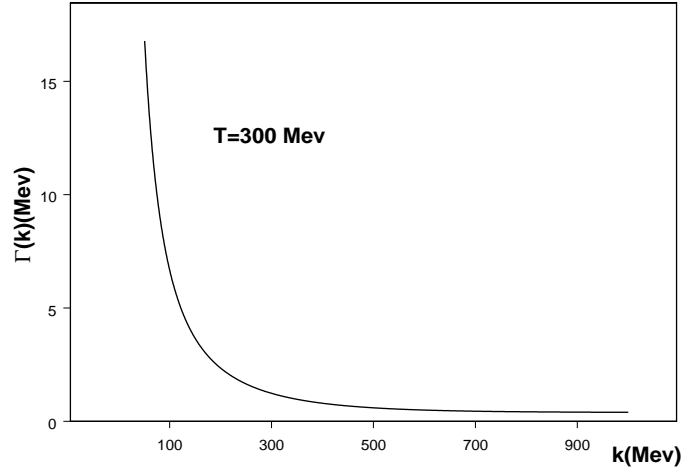,width=4in,height=3in}
 \caption{The approximation to the plasmon width given by (\ref{reswidth1})    for $\omega_{pl}=39.099~\mathrm{Mev}$, corresponding to a
 temperature $T= 300~\mathrm{Mev}$ .} \label{fig:gammak}
 \end{center}
\end{figure}

For $300\leq T \leq 450 ~\mbox{Mev}$ $\omega_p(k);\mathcal{Z}_k$
for $k \geq 200 ~\mbox{Mev}$ computed from equations
(\ref{imamodel}), (\ref{realsig1}) and (\ref{reswidth1}) are
numerically indistinguishable from those shown in figures
(\ref{fig:disprel},\ref{fig:Zofk}) respectively, while $\Gamma_k$
is displayed in fig. (\ref{fig:gammak}).

This figure in combination with the kinetic analysis of sections
(\ref{subsec:kinetics}) reveal an important aspect: the solution
of the kinetic equation (\ref{solkin}) shows that  the time scale
for thermalization of a photon with momentum $k$ is $t_{th;k} =
1/2\Gamma_k$. Figure (\ref{fig:gammak}) shows that for photons
with $k \leq 200 ~\mbox{Mev}$ the time scale for thermalization is
$\lesssim 15 ~\mbox{fm}/c$ which is of the same order as the
lifetime of the QGP expected at RHIC and LHC. Hence low energy
photons are expected to thermalize as \emph{quasiparticles} in a
QGP in LTE. This an important aspect that is completely missed by
the S-matrix approach which would assign these photons a constant
rate of emission even when they thermalize with the medium during
the lifetime of the QGP.

Below the light cone, the spectral density is determined by the
one loop Landau damping contribution and is given by eq.
(\ref{specphot}) with the real part of the photon self-energy
given by eq. (\ref{realsig1}) and the imaginary part given by
$\tilde{\Sigma}^{(1)}_{LD}(k;\omega)$ given by eq.
(\ref{imsigLD}).

In this approximation, the real time solution $f_k(t)$ is given by

\begin{eqnarray}\label{fofkfinal}
f_k(t)& = & \mathcal{Z}_k
\frac{\sin[\omega_p(k)\,t]}{\omega_p(k)}~e^{-\Gamma_k\, t}+ f_c(k,t)\nonumber \\
f_c(k,t) & = & \int_{-k}^{k} \frac{d\omega}{\pi} \frac{\sin(\omega
t)~\mathrm{Im}\tilde{\Sigma}^{(1)}_{LD}(k;\omega)}{\left[\omega^2-k^2-\mathrm{Re}\tilde{\Sigma}^{(1)}_T(k;\omega)
\right]^2+\left[\mathrm{Im}\tilde{\Sigma}^{(1)}_{LD}(k;\omega)\right]^2}
\end{eqnarray}

This solution represents a \emph{Dyson} resummation of the lowest
order contributions to the self-energy above and below the
lightcone within the approximations detailed above. The initial
condition $\dot{f}_c(k,0)=1$ is satisfied by the sum rule
(\ref{ressumrule}) which we confirmed numerically. Thus the
contribution from the Landau damping cut is necessary to fulfill
the initial condition and the sum rule.

With this real time solution, we now proceed to obtain the number
of photons emitted from a QGP in LTE up to a time $t$ by inserting
this solution eq. (\ref{fofkfinal}) in the expressions
(\ref{finumber}-\ref{funF}). Since we are only computing the
finite temperature contributions to the photon self-energy and
therefore neglecting the vacuum terms, we consistently set $Z=1$
in eq. (\ref{finumber}). As discussed in detail in
ref.\cite{nuestro} the zero temperature contribution to the
self-energy yields  the number of photons in the vacuum cloud
which is precisely cancelled by the \emph{vacuum} wave function
renormalization. Consistent with this result in ref.\cite{nuestro}
we neglect the zero temperature  contribution to the self-energy
and set $Z=1$, since both contributions cancel each other out in
the photon number\cite{nuestro}. Thus we focus solely on the
photons created in the medium. We have obtained the expression
(\ref{finumber}) under the assumption that the initial density
matrix is factorized and gaussian for photons. Of course this
assumption can be relaxed to contemplate initial states with
correlations between quarks and photons, however as discussed in
detail in ref.\cite{nuestro} such density matrix will \emph{not}
commute with $H_{QCD}$ since any entanglement between quark and
photon states will involve the quark current. Furthermore any
correlated  initial state will necessarily have to be modelled to
include the initial preparation during and after the collision as
discussed in detail in ref.\cite{nuestro}. In what follows we will
study the simplest assumption, namely that the initial state is
uncorrelated and with no photons. Accordingly, we set
$\mathcal{N}_k=0$ in eq. (\ref{finumber}). Introducing the
solution (\ref{fofkfinal}) into  eq. (\ref{finumber}) leads to
many different contributions. We will explicitly keep terms that
are of lowest order in $\alpha_{em}$ in the \emph{uniform}
expansion in the electromagnetic coupling. We emphasize a
\emph{uniform} expansion to contrast with a \emph{naive} expansion
which features terms that grow in time, namely secular terms. In
the many terms obtained we only keep those whose numerators are
manifestly of $\mathcal{O}(\alpha_{em})$ and neglect those which
are of higher orders. Specifically we find,

\begin{equation}
\frac{dN}{d^3x d^3k} = \frac{1}{(2\pi)^3}\sum_{i=1}^4 F_i(k,T,t)
\label{totalnumalfa}
\end{equation}
\noindent with the following contributions

\begin{eqnarray}
 F_1(k,T,t) & = & 2 n(\omega_p(k)) \mathcal{Z}_k(1-e^{-2\Gamma_k
 t}) \label{fac1} \\
 F_2(k,T,t) & = & 2\frac{e^{-2\Gamma_k  t}}{k}\int_{-k}^k
 \frac{d\omega}{\pi}\frac{\mathrm{Im}\tilde{\Sigma}^{(1)}_{LD}(k;\omega)}{\left[\omega
 - \omega_p(k)\right]^2+\Gamma^2_k}\left[1-\cos((\omega-\omega_p(k))t)
 \right]\label{fac2} \\
 F_3(k,T,t) & = & (\mathcal{Z}_k -1)(1-e^{-\Gamma_k
 t})^2 \label{fac3}\\
 F_4(k,T,t) & = & \frac{(1-e^{-\Gamma_k
 t})^2}{2k}\int_{-k}^k
 \frac{d\omega}{\pi}\frac{\mathrm{Im}\tilde{\Sigma}^{(1)}_{LD}(k;\omega)}{\left[\omega
 - \omega_p(k)\right]^2+\Gamma^2_k} (1+2n(\omega)) \label{fac4}
  \end{eqnarray}
\noindent with the wave function renormalization given to lowest
order by

\begin{equation}
\mathcal{Z}_k  =1-\int_{-k}^{k}
\frac{d\omega}{\pi}\frac{\omega~\mathrm{Im}\tilde{\Sigma}^{(1)}_{LD}(k;\omega)}{\left[\omega^2-k^2-\mathrm{Re}\tilde{\Sigma}^{(1)}_T(k;\omega)
\right]^2+\left[\mathrm{Im}\tilde{\Sigma}^{(1)}_{LD}(k;\omega)\right]^2}\label{ZLD}
\end{equation}

This expression summarizes the main result of this article: a
non-perturbative expression the determines the direct photon yield
directly in real time. It provides a Dyson-like resummation of the
naive perturbative expansion, it is uniformly valid at all times
and determines the yield to lowest order in $\alpha_{em}$.

A strict perturbative expansion of this result up to
$\mathcal{O}(\alpha_{em})$ leads to the following yield

\begin{equation}\label{pert}
\frac{dN_{PT}}{d^3x d^3k} = \frac{1}{(2\pi)^3}\left[ 2 n(k)
2\Gamma_k~t + \frac{2}{k}\int_{-k}^k
 \frac{d\omega}{\pi}\frac{\mathrm{Im}\tilde{\Sigma}^{(1)}_{LD}(k;\omega)}{\left[\omega
 - k\right]^2}\left[1-\cos((\omega-k)t)
 \right]\right]
\end{equation}

Taking the asymptotic long time limit in this expression, the
second term grows $\propto \ln(t)$ at long times\cite{nuestro}
hence it would be subleading and therefore neglected, while the
first term yields the S-matrix result for the rate

\begin{equation}\label{Smatxyield}
\frac{dN_{SM}}{d^4xd^3k} = 2n(k)(2\Gamma_k) = \frac{2~
\mathrm{Im}\tilde{\Sigma}^{PP}(k;\omega=k)}{(2\pi)^3~k(e^{\frac{k}{T}}-1)}
\end{equation}

\noindent with $\mathrm{Im}\tilde{\Sigma}^{PP}(k;\omega=k)$ given
by eq. (\ref{imsigPP}).

Thus it is clear that our final result for the photon yield during
a finite time $t$ given by eqns. (\ref{totalnumalfa}-\ref{fac4})
is a truly non-perturbative  resummation of the naive perturbative
expansion that incorporates real-time processes not captured by
the S-matrix approach.

Furthermore this result clearly reveals several physical aspects
that   are missed by the S-matrix approach and that underlie its
  shortcomings:

\begin{itemize}

\item{   In the asymptotic long time limit, the only terms
  that survive are (\ref{fac1},\ref{fac3},\ref{fac4}) which in the perturbative limit
 $\mathcal{Z}_k=   1+\mathcal{O}(\alpha_{em})$,
  reveal the correct   physics: in a QGP in equilibrium with infinite lifetime, photons
  will thermalize as \emph{quasiparticles} in the medium. Their
  abundance is determined by the Bose-Einstein distribution function for
  the correct quasiparticle frequency,  and the wave function
  renormalization measures the weight of the quasiparticle in the
  spectral density.}

\item{The width $\Gamma_k$ is formally of
$\mathcal{O}(\alpha_{em})$, however expanding the expressions
(\ref{fac1}-\ref{fac4}) in $\Gamma_k$, namely $e^{-\Gamma_k t}
\sim 1-\Gamma_k t + \frac{1}{2}\Gamma^2_k t^2
+\mathcal{O}(\Gamma^3_k t^3) $ will lead to \emph{secular} terms
in time, namely terms that grow in time \emph{to all orders in
perturbation theory}. Similarly, since $\omega_p(k)-k$ is formally
of $\mathcal{O}(\alpha_{em})$ one would be tempted to expand the
argument of the cosine in eq. (\ref{fac2}), this again will lead
to secular terms in time. Finally a naive perturbative expansion
will attempt to replace $\omega_p(k) \rightarrow k, \Gamma_k
\rightarrow 0$ in the denominators of eq. (\ref{fac2}) and
(\ref{fac4}), however this will lead to logarithmic divergences at
the Landau damping threshold which manifest themselves as
logarithmic secular terms in time\cite{nuestro}. The S-matrix
approach computes the transition probability per unit space-time
volume  by taking the infinite time limit and extracts the linear
term in time which gives a constant rate. Obviously taking the
long time limit is manifestly ignoring the fact that however long
the photon relaxation time, eventually photons will thermalize in
the long time limit. A \emph{kinetic} interpretation of the
S-matrix result, as discussed in section (\ref{subsec:kinetics})
is given by

\begin{equation}
\frac{dN_{SM}}{d^4xd^3k} = \left.
\frac{dN_{kin}}{dtd^3xd^3k}\right|_{t=0}
\end{equation}

\noindent setting $\omega_p(k) \rightarrow k, \mathcal{Z}_k
\rightarrow 1$ to lowest order in $\mathcal{O}(\alpha_{em})$, thus
obtaining the result given by eq. (\ref{rateSm}) of section
(\ref{subsec:kinetics}).

The non-perturbative aspects of the main result of this article,
namely equations (\ref{totalnumalfa}-\ref{fac4}) make manifestly
clear the shortcomings of the S-matrix approach to extract the
photon emission yield, including those in ref.\cite{serreau}: such
approach treats the plasma as a state in LTE of infinite lifetime
and assumes that the photons are produced at a constant rate.
These assumptions manifestly ignore the physically correct picture
that if the lifetime of the QGP is infinite then photons produced
will eventually thermalize. The validity of the S-matrix approach
will then have to be justified on some intermediate time scale $t
\ll 1/\Gamma_k$ but then finite time effects \emph{must} be
included in the full calculation and the validity of the S-matrix
approach will depend on the momentum of the photon. As we have
discussed  above, long-wavelength photons with $k \leq
200~\mbox{Mev}$  thermalize on time scales of the order of the
expected lifetime of the QGP at RHIC and LHC $t \sim 10-20
~\mbox{fm}/c$ (see fig. (\ref{fig:gammak})), thus the S-matrix
calculation is explicitly  restricted to large momentum photons
and only during a time scale $t \ll 1/\Gamma_k$ which depends on
the momentum. But clearly this is in manifest contradiction with
the assumptions underlying the S-matrix calculation: the existence
of asymptotic states and taking the infinite time limit.

Assuming an initial state prepared in the asymptotic past not only
ignores the true physical aspects of the problem, namely that
quarks and gluons are partonic degrees of freedom in the incoming
colliding hadrons, not \emph{asymptotic states}, but also that
however long the mean-free path of the photon, if the plasma lives
long enough, these will thermalize.  }

\item{ In obtaining the final result given by equations
(\ref{totalnumalfa}-\ref{fac4}) we have systematically neglected
terms that are truly perturbative and of higher order in an
uniform expansion in $\alpha_{em}$. Namely the terms that we
neglected are all bound in time and manifestly of higher order in
$\alpha_{em}$. The final result (\ref{totalnumalfa}-\ref{fac4})
provides a resummation of the naive perturbative expansion much in
the same manner as the dynamical renormalization
group\cite{boyankinetic} which leads to correct kinetic
equations\cite{boyanphoton}.}

  \end{itemize}

Our main results and predictions for the yield and spectrum of the
emitted photons are displayed below in figure
(\ref{fig:lognumber}).

This figure shows the yield obtained from the result eqs.
(\ref{totalnumalfa}-\ref{fac4}) as compared to the S-matrix yield
for $T=300~\mbox{Mev}; t=10~\mbox{fm}/c$ which are
 values of the temperature and lifetime expected at RHIC. We have
 displayed the results only up to $k \sim 5~\mbox{Gev}$ since as
 discussed in \cite{nuestro} and above, for much larger momenta
 the wavelength of the photon is likely probing scales shorter than the
 mean free path and the approximation of LTE is no longer
 reliable. Furthermore, as discussed in detail in \cite{nuestro},
 for large momenta $k \geq 4-5 ~\mbox{Gev}$ the spectrum of
 photons is sensitive to the initial pre-equilibrium stage\cite{renk}.

\begin{figure}[htbp]
\begin{center}
\epsfig{file=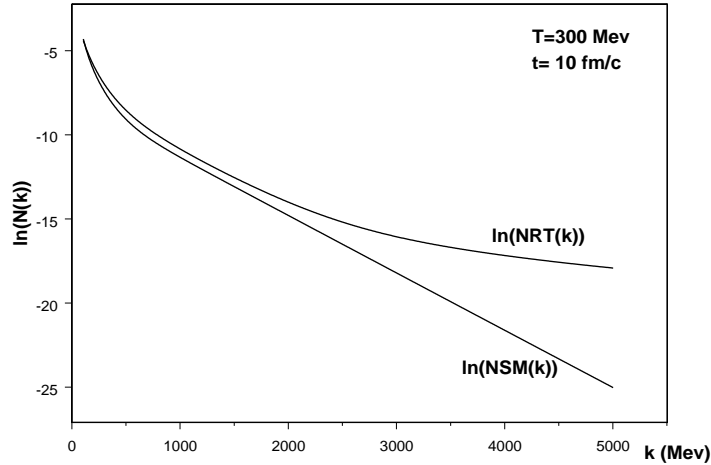,width=4in,height=3in}
 \caption{ $\ln\left(\frac{dN}{d^3k d^3 x}\right)$ vs. $k(\mbox{Mev})$ for $T=300~\mbox{Mev}$ and $t=10~\mbox{fm}/c$.
  $NRT(k)=dN/d^3kd^3x$ given by eqs.
(\ref{totalnumalfa}-\ref{fac4}) and $NSM(k)= dN_{SM}/d^3kd^3x$
given by  eqs. (\ref{Smatxyield},\ref{imsigPP}).  }
\label{fig:lognumber}
 \end{center}
\end{figure}

Clearly the real time yield is systematically larger than that
predicted by the S-matrix approach. A flattening of the spectrum
at $k \sim 2.5-3 ~\mbox{Gev}$ is a distinct feature that
originates in processes that while subleading in the
asymptotically long time limit  dominate during the finite
lifetime of the QGP. We have established numerically that amongst
the four terms in eqs. (\ref{fac1}-\ref{fac4}), the largest
contributions to the yield during the QGP lifetime  $\sim
10~\mbox{fm}/c$ are those from eq. (\ref{fac1}) and eq.
(\ref{fac2}) while the contributions from eq.(\ref{fac3}),
(\ref{fac4}) are numerically smaller for the range of momenta
displayed in fig. (\ref{fig:lognumber}). The first term given by
(\ref{fac1}) features the fastest fall off with energy, while the
remaining terms (\ref{fac2}-\ref{fac4}) all contribute to the
flattening of the spectrum.
\section{Discussion:}
\subsection{Time evolution in URHIC:} An important aspect of the
real time formulation of photon production is the realization that
particle production during ultrarelativistic heavy ion collisions
requires the understanding of dynamics \emph{very different from
usual collision experiments}. In usual collision experiments the
colliding particles in the initial beam are the `in' states and the
reaction products are the `out' states. In and out states are
eigenstates of the non-interacting Hamiltonian with the physical
masses. The overlap between these states and those created by the
Heisenberg field operators  out of the exact vacuum at
asymptotically early and late times is determined by the
\emph{vacuum}  wavefunction renormalization constant $Z$.

Time evolution in quantum mechanics and quantum field theory is an
\emph{initial value problem} and this is the basic ingredient in
S-matrix theory approach to scattering. The S-matrix approach
considers the time evolution of initial states corresponding to the
`in' states, namely eigenstates of the free Hamiltonian with the
physical masses which are prepared in the infinite past. The
S-matrix is the time evolution operator (in the interaction picture)
from $t_i=-\infty$ up to $t_f=\infty$ and evolves these initial
states up to an asymptotically large time, at which the overlap
between the `out' states, eigenstates of the free field Hamiltonian
of physical mass is computed. The usual reduction formulas introduce
the factor $Z$ to account for the overlap between the states created
by the full Heisenberg field operator and those of the free field
creation operators of in and out fields.

In URHIC the initial states are heavy ions and the final states are
mesons and baryons, not the elementary quark and gluon fields in the
QCD Hamiltonian.  The collision involves the deconfinement,
thermalization, and subsequent re-hadronization of partons. The
correct S-matrix approach should then include \emph{all aspects} of
these dynamical processes in order to provide a consistent framework
for calculation. Needless to say, this is currently an impossible
task. Instead, previous approaches to photon production seek to
obtain the photon yield from the \emph{different stages
independently}. The pre-equilibrium stage is studied via parton
cascade models which input parton scattering cross sections computed
from S-matrix theory into a semiclassical transport formulation. The
thermalized QGP state is studied by assuming a stationary QGP of
infinite lifetime in thermal equilibrium with no memory of the
initial pre-equilibrium stage. Photons from the hadronized state are
studied by assuming a `cocktail' of mesons. The results from
S-matrix computations (which necessarily \emph{assume} an infinite
lifetime) in a thermalized QGP are combined with hydrodynamics to
integrate the \emph{rate} during the finite lifetime of the QGP,
manifestly ignoring that the S-matrix approach has assumed an
infinite lifetime. Photon production during the hadronized stage is
calculated via S-matrix with in and out states corresponding to
(short lived) mesons. Finally the total spectrum of photons is
obtained by adding the contributions from the different stages. It
should be obvious that there are glaring shortcomings in this
program which justify deeper scrutiny.

The real-time approach that we advocate is solidly based on the
basic premise of quantum mechanics and quantum field theory, namely
that time evolution is an initial value problem: given an initial
state (or density matrix) and the  Hamiltonian, time evolution is
completely determined. Basic quantum mechanics states that given a
density matrix at some initial time $t_i$, $\widehat{\rho}(t_i)$,
the density matrix at \emph{any} time $t$ (in the Schroedinger
picture) is \be \widehat{\rho}(t)= e^{-iH(t-t_i)} \;
\widehat{\rho}(t_i) \; e^{iH(t-t_i)} \, , \ee \noindent where $H$ is
the complete Hamiltonian. In the case under consideration, $H$ is
the full Hamiltonian of QCD plus electromagnetism and is \emph{time
independent}. It is convenient to pass to the interaction picture of
the non-interacting Hamiltonian $H_0$ in which the evolution
operator is given by the identity \be\label{opi} e^{-iH(t-t_i)} =
e^{-iH_0\,t} \;  U(t,t_i) \;  e^{iH_0\,t_i} \ee \noindent where
$U(t,t_i)$ is the usual unitary time evolution operator in the
interaction picture.

The expectation value of \emph{any} arbitrary Schr\"odinger picture
operator is \be \label{RTfots} \langle \mathcal{O}\rangle(t) =
\frac{\mathrm{Tr} \, \left[ \mathcal{O}\widehat{\rho}(t)
\right]}{\mathrm{Tr}\, \widehat{\rho}(t_i)} \ee This is the same as
the expectation value of the Heisenberg picture operator in the
initial density matrix. Computing real time expectation values is
\emph{very different} from computing S-matrix elements. The former
are in-in matrix elements as befits time evolution as an initial
value problem, the latter gives a transition amplitude from a state
far in the past to a state far in the future and is conceptually
very different from obtaining an expectation value of an operator in
a time evolved state.

It remains to define the Heisenberg number operator: the Heisenberg
\emph{field} operator for photons creates a single photon state (in
or out) out of the \emph{vacuum} with probability $\sqrt{Z}$, with
$Z$ being the on-shell vacuum wave function renormalization
constant. Since the number operator is a bilineal in the field it
must be divided by $Z$ in order to ensure that it counts asymptotic
photon states with weight one, this is an important ingredient of
the reduction formulae. Furthermore in free field theory a simple
normal ordering suffices to ensure that the number operator
annihilates the (bare) vacuum. However in an interacting theory, a
normal ordering constant $\mathcal{C}_k$ \emph{must} be introduced
so that the expectation value of the Heisenberg field operator
vanishes in the exact vacuum. This normal ordering constant can be
simply extracted from the equal time limit of the operator product
expansion (OPE) of the correlation function of the photon field. The
short distance expansion is completely determined by the OPE in the
vacuum. Therefore subtracting the constant $\mathcal{C}_k$ is
\emph{necessary} to define a normal ordered expectation value of the
Heisenberg number operator consistently with the equal time limit of
the OPE of the correlation function of the (transverse) gauge field.

This is \emph{precisely} the main point of our program, we obtain
the expectation value of the photon number operator given by eq.
(\ref{numberop}) which includes the overlap factor $Z$ and the
normal ordering subtraction $\mathcal{C}_k$ \emph{both} determined
in the \emph{vacuum} consistently with asymptotic theory. In this
article we implemented this program by obtaining the non-equilibrium
effective action for the gauge field as explained in section
(\ref{sec:action}). As was explicitly shown in our previous article
(see \cite{nuestro}), the vacuum wave function renormalization
constant and the vacuum subtraction $\mathcal{C}_k$ cancel the
vacuum contributions to the number operator. That this is indeed the
number of photons (exact massless eigenstates of the free photon
Hamiltonian) produced   in the medium up to time $t$ by the
evolution is clear from the following trivial equality \be
\label{form} \langle N_k\rangle (t) = \langle N_k\rangle (t_i)+
\int^t_{t_i} \frac{d\langle N_k\rangle (t')}{dt'} \ee \emph{If} the
photons do not undergo further interactions after being produced
during the lifetime of the QGP, they will propagate freely towards
the detector.

As has been shown explicitly in section (\ref{loword}) the S-matrix
(or alternatively the kinetic approach) extracts the rate
$d<N_k>(t)/dt$ from Fermi's Golden rule, by taking the time interval
to infinity which leads to a time independent rate, and inputs this
rate in the calculation of the yield just as in eq. (\ref{form})
above accounting for hydrodynamical expansion.

Of course if one \emph{knew} how to evolve an initial heavy ion
state through deconfinement, thermalization and finally
re-hadronization with the full QCD Hamiltonian, one could begin with
an initial density matrix of heavy ions at $t_i\rightarrow -\infty$
and compute the total number of photons produced \emph{all
throughout the dynamics} up to $t\rightarrow +\infty$ by
implementing (\ref{RTfots}). Obviously this program is not feasible.
However, the statement of formation of a thermalized QGP at a time
of order $\sim 1~\mathrm{fm}/c$ suggests that at this \emph{initial
time} $t_i$ the density matrix is that of thermal equilibrium for
the QGP. Time evolution being an initial value problem, \emph{if}
the initial density matrix is \emph{completely} specified, then we
can evolve it in time with the full \emph{time independent}
Hamiltonian $H$. Perturbation theory will be reliable if quarks and
gluons are weakly interacting, the usual assumption for a QGT in LTE
but certainly \emph{not} during hadronization. Nevertheless, we can
still obtain the number of photons produced up to any arbitrary time
$t$ from a weakly interacting QGP in LTE  from eq. (\ref{RTfots})
\emph{if} we knew exactly what is the distribution of photons in the
initial state. The main uncertainty in the calculation of the yield
is precisely the knowledge of the photon distribution function in
the initial density matrix.

Only when photons reach thermal equilibrium with the medium will
their spectrum and distribution be \emph{independent of the initial
conditions}. However if photons do not equilibrate, and this is the
usual  assumption, the spectrum of produced photons \emph{will}
depend on the initial state. In ref.\cite{nuestro} we have studied
suitably modeled initial states and found that the photon spectrum
is sensitive to the details of the initial state for energy larger
than $\sim 5-6 \mathrm{Gev}$ assuming a thermalization scale
$\lesssim 1\mathrm{fm}/c$.

\vspace{1mm}

Recently a criticism has been leveled at the real time formulation
for studying photon production from a transient QGP\cite{schiff}
that we propose. The authors in this reference claim that the
unitary time evolution operator $U(t,t_i)$ in the interaction
picture in eq. (\ref{opi}) corresponds to `switching on' the
interaction at $t=t_i$ and `switching-off' at time $t_f$. This is
clearly incorrect as the identity Eq. (\ref{opi}) makes manifest.
Such a switch-on-off of the interaction would entail a time
\emph{dependent} Hamiltonian with an artificial time dependent
coupling. The identity (\ref{opi}) clearly states that time
evolution is obtained with the full time independent Hamiltonian of
QCD plus QED. This statement by the authors of ref.\cite{schiff}
reflects a misunderstanding  of the basic fact that time evolution
in quantum mechanics and quantum field theory is an \emph{initial
value problem}: a state specified at some initial time $t_i$ is
evolved in time up to an arbitrary time $t_f$ with the unitary time
evolution operator given by eq. (\ref{opi}). As we mentioned above
once the density matrix (or a pure state) is specified at \emph{any}
arbitrary time $t_i$ it can be evolved up to another arbitrary time
$t$ with the unitary time evolution operator. Any statements about
`switching on-off' of interactions is simply a misunderstanding of
this basic fact in quantum mechanics. A series of formal steps are
invoked in ref.\cite{schiff} to incorrectly argue that only taking
$t_f\rightarrow \infty$ gives the total photon number. With this
statement the authors are ignoring or glossing over a large body of
work in the hydrodynamic approach to photon production from a
thermalized QGP. In this approach an invariant rate, obtained
assuming an infinite lifetime is input in the hydrodynamical
evolution, and the total photon yield is obtained by integrating
over the space-time history of the QGP, namely over a \emph{finite
lifetime}. One of our main criticisms of this approach is that the
rate input in these calculations is obtained from an S-matrix
computation which \emph{assumes} an infinite lifetime, but is used
during a finite lifetime.

In a plasma that expands nearly at the speed of light, if its
lifetime is infinite as the authors suggest, photons will ultimately
thermalize with the plasma. Physically the QGP only `lives' for a
few $\mathrm{fm}/c$, furthermore as the plasma expands and cools the
coupling grows stronger and the theory becomes strongly coupled at
the hadronization phase transition. Any formulation that attempts to
use perturbation theory in a long time limit is obviously flawed.
All of these physical arguments point out the limitations of an
S-matrix approach and the inconsistency in taking an infinite time
limit.

The authors in ref.\cite{schiff} completely ignore the physical
arguments that apply to the experimental situation therefore either
ignoring or glossing over the inconsistency in taking an infinite
time interval. Nor do they address the important issue of what is
the initial state when the QGP is conjectured to thermalize, in
particular what is the initial photon distribution function.
S-matrix theory simply takes the initial time $t_i\rightarrow
-\infty$ and takes the initial states to be free field `in' states.
Clearly this is not the physical situation in a heavy ion collision
and the pre-equilibrium stage is important.

In ref. \cite{schiff} the discussion stops short of actually
calculating the number of photons in their limit $t_f\rightarrow
\infty$. Had they done so they would have found a result that
diverges linearly with $t_f$. This is precisely the result of the
S-matrix calculation where $t_f$ is interpreted as the total
`reaction time' and the result is divided by $t_f$ to yield the
rate. At this stage the physical question that the authors did not
ask themselves is what is this `reaction time'. Obviously if these
photons are produced by a transient QGP, $t_f$ must be the lifetime
of this state. Lest that the authors actually mean to follow the
dynamics through the confinement and hadronization phase transition,
in which case their perturbative arguments are naive at best.

The authors also question our definition of the Heisenberg number
operator that includes the vacuum wave function renormalization $Z$
and normal ordering constant $\mathcal{C}$. As explained in the text
and again explicitly in this section above, the wave function
renormalization is required because the interpolating Heisenberg
operator creates a single in or out photon state with amplitude
$\sqrt{Z}$ out of the vacuum. This is a basic result of asymptotic
theory and the factor $1/Z$ is the usual factor that accompanies all
bilinear correlation functions (such as the Green's function) in the
reduction formulae, this is a basic issue in renormalization and in
the LSZ formulation (see for example\cite{BD}). The normal ordering
constant is a simple statement that the equal time limit in the
operator product expansion gives a non-vanishing expectation value
in the \emph{vacuum} and must be subtracted. This is the basic
statement that normal ordering and time evolution \emph{do not
commute} in an interacting field theory. Even in free field theory
the number operator is defined with a normal ordering prescription
(the usual symmetrization factor $1/2$). Such normal ordering must
be re-defined in the interacting theory and that is precisely what
$\mathcal{C}$ achieves. It is defined at zero temperature, namely in
the vacuum and as explicitly shown in ref.\cite{nuestro} it cancels
the vacuum contributions to the photon number in the interacting
theory.

In summary, the criticisms leveled in ref.\cite{schiff} have no
credibility: i) statements such as `switching-on' and `off' reflect
a misunderstanding of basic quantum mechanics and quantum field
theory, ii) formal statements about infinite time limits etc, are
only glossing over the \emph{physical} situation under
consideration, namely a transient state which undergoes a non-
perturbative phase transition at a finite time, iii) the statements
regarding wave function renormalization and subtractions also
reflect a misunderstanding of basic renormalization issues in
quantum field theory, iv) the authors do not even attempt to discuss
the issue of the initial state, v) the criticisms in ref.
\cite{schiff} do not attempt to scrutinize the usual assumptions
with their glaring inconsistencies, nor do they provide a reliable
framework to study the \emph{physical situation} at hand,  vi) the
non-perturbative aspects associated with the \emph{propagation} of
photons in the medium (see discussion below) are not included in
ref. \cite{schiff}, which our formulation in terms of the real time
effective action describes consistently and systematically. Hence,
these criticisms not only do not help to understand (or even
address) the important and very relevant physical questions but
disguise these fundamental aspects of URHIC physics.

\subsection{Non-perturbative aspects:}
The formulation of photon production based on the real-time
effective action presented in this article allows to understand the
following \emph{non-perturbative aspects} of propagation and
relaxation.

  {\bf i) Propagation:} The photons produced in a QGP in LTE do not propagate
as free waves in
 vacuum but  \emph{in a medium} with an `index of
 refraction'. The propagation aspects of photons in this
 \emph{medium} require that self-energy corrections to the
 propagator be included. A simple perturbative expansion of the
 propagator in terms of the  photon self-energy $\Sigma(\omega,k)$
 is
 \be G(\omega,k)=
 \frac{1}{\omega^2-k^2}+\frac{1}{\omega^2-k^2}\,\Sigma(\omega,k)\,
\frac{1}{\omega^2-k^2}+\cdots
 \ee
 \noindent obviously features divergences on the photon mass shell,
 which in real time lead to terms that grow  in
 time as for example featured in eq. (\ref{pert}). A Dyson (geometric)
resummation of these self-energy
 corrections leads to the propagator
\be \label{Dyson} G(\omega,k)=
 \frac{1}{\omega^2-k^2-\Sigma(\omega,k)}
\ee
 The poles in this resummed propagator describe the correct modes of
 propagation in the medium, namely the transverse plasmons and their
relaxational dynamics. As usual
 the self-energy $\Sigma(\omega,k)$ is calculated in perturbation
 theory up to a given order in the coupling, but the Dyson resummed
 propagator eq. (\ref{Dyson}) provides an \emph{all order resummation of
 select diagrams}. It is precisely in this sense that the real-time
 effective action leads to a non-perturbative description which
 describes the correct propagation of photons in the medium as plasmon modes.
In section \ref{sec:plasmon} we discussed in detail the main aspects
of photon
 propagation gleaned from the hard-thermal loop approximation to the
 self-energy. The poles in the Dyson resummed propagator determine
 the propagating modes, their frequencies and widths. Of course the
 dispersion relation and widths are obtained in perturbation theory
 up to the order in which the self-energy has been computed. This is
 the usual statement in many body or field theory that in order to
 obtain the position of the poles (propagating modes) a Dyson
 resummation of the self-energy corrections must be performed.
 These corrections that account for the propagation of photons in
 the medium with the correct dispersion relation are featured in the
 final equations (\ref{totalnumalfa})-(\ref{fac4}) in the wave
 function renormalization $\mathcal{Z}_k$, the dispersion relation
 $\omega_p(k)$ and the width $\Gamma_k$ in the denominators.

{\bf ii) Relaxation:} Our final equation for the yield of photons
produced during the lifetime of the QGP
(\ref{totalnumalfa})-(\ref{fac4}) features contributions with the
factor $1-e^{-\Gamma_k t}$ with $\Gamma_k$ of
$\mathcal{O}(\alpha_{em})$ and up to leading logarithms in
$\alpha_s$. The naive perturbative expansion in $\alpha_{em}$
replaces these factors by $ 1-e^{-\Gamma_k t}  \sim \Gamma_k t $,
which is the usual yield obtained from the S-matrix calculation.
However, such approximation is obviously only valid for $\Gamma_k t
\ll 1$ (which again makes manifest the shortcoming of the S-matrix
approach). However, as studied in detail in section \ref{beyo}
photons of energy $\sim 200 \mathrm{Mev}$ thermalize in the QGP
during its lifetime of about $10 \mathrm{fm}/c$ and the product
$\Gamma_k t$ is a sizable fraction of unity for photons up to
energies $\lesssim 500-700 \mathrm{Mev}$. Namely the full solution
given by eqs.(\ref{totalnumalfa})-(\ref{fac4}) accounts for the
partial thermalization of medium energy photons as a result of the
non-perturbative Dyson resummation and is explicit in the
exponentials which result from the complex poles in the Dyson
resummed propagators. Naive perturbation theory certainly misses
thermalization since the yield obtained from a perturbative S-matrix
calculation will only feature a term of the form $\Gamma_k \, t$ and
it also misses the correct propagation of photons in the medium.

\section{Conclusions, summary and further questions:}

In this article we have focused on studying  production of photons
from a QGP in LTE by implementing a non-perturbative formulation
of the real-time evolution of an initial density matrix. The main
ingredient is the real time effective action for the
electromagnetic field \emph{exact} to order $\alpha_{em}$ and in
principle to all orders in $\alpha_s$. The real time evolution is
completely determined by the solution of a non-local stochastic
Langevin equation which results in a Dyson-like resummation of the
naive perturbative expansion of the photon self-energy. A quantum
kinetic description emerges directly from this non-perturbative
formulation.

The main result for the direct photon yield from a QGP in LTE is
given by eqns. (\ref{finumber}) with (\ref{foft},\ref{flucdiss}).
We have confirmed that in a strict perturbative expansion, eq.
(\ref{finumber}) reproduces the S-matrix results as well as
previously obtained results with initially uncorrelated
states\cite{wangphoton,nuestro}. Furthermore we have shown that
the result (\ref{finumber}) reproduces the main features of
quantum kinetics for the evolution of the photon distribution
function.

An explicit result for the photon self-energy to order
$\alpha_{em}$ and up to leading logarithmic order in
$\alpha_s$\cite{AMY} indicates that photons with momenta $k
\lesssim 200~\mbox{Mev}$ propagate as plasmon quasiparticles and
\emph{thermalize} on a time scale $t \leq 10-15 ~\mbox{fm}/c$,
namely of the order of the lifetime of the QGP expected at RHIC
and LHC.

To leading order in a \emph{uniform } expansion in $\alpha_{em}$
and up to leading logarithmic order in $\alpha_s$ our main result
is  summarized by equations (\ref{totalnumalfa}-\ref{fac4}) and
our  prediction for the spectrum of photons from a QGP in LTE with
a lifetime of order $\sim 10~\mbox{fm}/c$  is displayed in fig.
(\ref{fig:lognumber}).

In the energy range of experimental relevance which we deem
theoretically trustworthy $200 ~\mbox{Mev} \lesssim k \lesssim 4-5
~\mbox{Gev}$ the real time yield obtained in leading order and
given by equations (\ref{totalnumalfa}-\ref{fac4}) is displayed in
fig. (\ref{fig:lognumber}). It is systematically \emph{larger}
than previous estimates based on the S-matrix approach and
features a flattening of the spectrum for $k \geq 2.5 -3
~\mbox{Gev}$. We conclude that during the finite lifetime of a QGP
in LTE expected at RHIC and LHC there are important real time
processes that contribute to photon emission that are not captured
by the usual approach. Furthermore, a distinct prediction
resulting from our study is a flattening of the spectrum, a
feature that is associated precisely with these real time
processes.

We have also discussed the limit of reliability of our results and
the theoretical uncertainties that are present in \emph{any}
formulation of photon emission from a QGP in LTE: i) theoretical
uncertainties in the initial state and pre-equilibrium stage,
these will manifest themselves in the high energy part of the
spectrum\cite{renk,nuestro}, ii) a lack of knowledge of the
imaginary part of the photon self-energy for \emph{all } values of
the frequency, iii) a likely breakdown of the reliability  of the
assumption of LTE for large energy photons. Once these
uncertainties are resolved, a knowledge of the initial state
(which itself requires understanding of the pre-equilibrium stage)
and of the imaginary part of the photon self-energy can be
systematically input in our non-perturbative formulation.

 \acknowledgements  D. B.  thanks the N.S.F. for
partial support through grant PHY-0242134, and the hospitality of
LPTHE and the Observatoire de Paris where part of this work was
carried out. He also thanks P. Bedaque, X.-N. Wang, V. Koch, J.
Randrup and A. Majumder for  illuminating discussions and the
Nuclear Theory group at Lawrence Berkeley National Lab. for
hospitality during part of this work.  H. J. d. V. thanks the
Department of Physics and Astronomy, University of Pittsburgh for
their hospitality.

\appendix

\end{document}